\documentclass[pra,twocolumn,showpacs,preprintnumbers,superscriptaddress]{revtex4}
\usepackage{amsmath,amssymb,mathrsfs}
\usepackage{subfigure}
\usepackage[dvips]{graphicx}
\usepackage{hyperref}
\usepackage{paralist}
\usepackage{epstopdf}
\usepackage{dcolumn}
\usepackage{color}

\def\be{\begin{equation}}
\def\ee{\end{equation}}
\def\bea{\begin{eqnarray}}
\def\eea{\end{eqnarray}}

\def\tbf{\textbf}
\newcounter{RomanNumber}
\newcommand{\MyRoman}[1]{\setcounter{RomanNumber}{#1}\Roman{RomanNumber}}

\begin{document}

\title{Effective action approach to the p-band Mott insulator and superfluid transition}
\author{Xiaopeng Li}
\affiliation{Department of Physics and Astronomy, University of Pittsburgh, Pittsburgh, Pennsylvania 15260, USA}
\affiliation{Kavli Institute for Theoretical Physics, University of
  California, Santa Barbara, CA 93106, USA}

\author{Erhai Zhao}
\affiliation{Department of Physics and Astronomy, George Mason University,  Fairfax, Virginia 22030, USA}

\author{W. Vincent Liu}
\email[e-mail:]{w.vincent.liu@gmail.com}
\affiliation{Department of Physics and Astronomy, University of
Pittsburgh, Pittsburgh, Pennsylvania 15260, USA}
\affiliation{Kavli Institute for Theoretical Physics, University of
  California, Santa Barbara, CA 93106, USA}

\begin{abstract}
{ Motivated by the recent experiment on p-orbital band bosons in optical lattices, we study theoretically
the quantum phases of Mott insulator and superfluidity in two-dimensions.  The system features a novel superfluid phase with transversely staggered orbital current
at weak interaction, and a Mott insulator phase with antiferro-orbital order at strong coupling
and commensurate filling. We go beyond mean field theory and derive from a
microscopic model an effective action that is capable of describing both the p-band Mott insulating and  superfluid phases in strong coupling. We further calculate the excitation spectra near the quantum critical
point and find two gapless modes away from the tip of the Mott lobe but four gapless modes at
the tip. Our effective theory reveals how the phase coherence peak builds up in the Mott regime
when approaching the critical point. We also discuss the finite temperature phase transition of
p-band superfluidity.
}

\end{abstract}
\preprint{NSF-KITP-11-041}
\pacs{03.75.-b, 05.30.Rt, 64.60.My, 67.85.-d}
\maketitle

\section{Introduction}
{ Recent years have witnessed a dramatic experimental progress on cold atom systems in optical lattices~\cite{Bloch_review,Lewenstein_review}}. Detailed studies of strongly correlated quantum systems are possible thanks to the unprecedented controllability of cold atom experiments. The lowest band Mott-superfluid transition~\cite{Fisher_bh,Jaksch_mottsf} for lattice bosons has been successfully observed in experiments~\cite{Greiner_mottsf}. Experiments on populating bosons on excited bands are also put forward~\cite{Browaeys_pband,Muller_pband}. And a more recent p-band boson experiment~\cite{Lewenstein-Liu_NPhysnews,Wirth_pband} {opens up a new thrust towards observing the exotic phases of bosons on higher bands with long life time.}  Given growing experimental progress on excited band bosons, probing detailed features of p-band Mott insulators and p-band superfluidity {is attracting broader interests.} There are numerous theoretical work on excited band bosons focusing on proposing exotic phases~\cite{Isacsson_pband,Liu_TSOC,Wu_pband,Kuklov_sf,Lim_TSOC}, which have demonstrated the fascinating physics associated with bosons on excited bands of optical lattices. {The quantum phase transition from p-band Mott insulator to superfluid phase has been studied within the Gutzwiller mean field approach~\cite{Isacsson_pband,Collin_pband}.}
%This transition, due to competition between kinetic energy and repulsive on-site interactions between p band bosons is driven by quantum fluctuations at zero temperature.
For large interactions, the energy is minimized by an incompressible state with an orbital order. And for weak interactions the kinetic energy dominates over the interaction and drives the system into a superfluid with a feature of transversely staggered orbital current (TSOC)~\cite{Liu_TSOC}. The competition between the kinetic energy and interaction energy is well described within the Gutzwiller approach, however the single particle spectra and the momentum distribution are out of reach within this approach.

{In this paper we  apply the method of effective action beyond the Gutzwiller mean field,} %propose a theory beyond the Gutzwiller mean field approach,
and explore the single particle spectra in both of the p-band Mott insulator phase and the TSOC superfluid phase in two dimensions. {We have studied the phase coherence in the Mott insulator phase and found that sharp peaks rise at finite momenta (($\pm \pi, 0$) for the p$_x$ band and ($0, \pm \pi$) for the p$_y$ band) when the Mott gap is small.} This offers new approaches of preparing coherent matter waves from Mott insulators. From the p-band Mott insulator phase to the TSOC superfluid phase, the global U(1) symmetry and the time reversal T symmetry are broken. {Away from the tip of the Mott lobe, %the Mott tip regime,
we find two gapless modes at the critical point; while at the tip, we find four gapless modes due to the particle-hole symmetry.} {For TSOC superfluid phase we go beyond previous study %on the single particle spectra of TSOC superfluid phase
in the weak coupling limit~\cite{Liu_TSOC,Wu_pband} and consider the leading effect of Hubbard interaction in the strong coupling regime. Our theory is capable of capturing the main feature of TSOC superfluid phase in the strong coupling regime, where the critical point of Mott-superfluid transition is located. The isotropy of the sound velocity of the TSOC superfluid phase is explained. Finally, the finite temperature phase transitions of the strongly interacting TSOC superfluid phase are discussed.}
%{\bf check every point is shown in the paper in an apparent way.}

The paper is organized as follows. In Sec. \MyRoman{2}, we discuss the microscopic model describing the extended Bose-Hubbard model with p-orbital degrees of freedom and revisit the features of the p-band Mott insulator phase and the TSOC superfluid phase. In Sec. \MyRoman{3}, we derive an effective action for the extended Bose-Hubbard model in the strong-coupling limit by performing two successive Hubbard-Stratonovich transformations of the inter-site hopping term. {In Sec. \MyRoman{4},  we calculate the single particle spectra and the momentum distribution for the p-band Mott insulator phase.
We apply Bogoliubov theory on our effective action and study the sound velocity in the TSOC superfluid phase.} The finite temperature phase transitions of TSOC superfluid phase are also discussed in Sec. \MyRoman{4}. We conclude with a brief discussion in Sec. \MyRoman{5}.

\section{Model and phase diagram}
We start with a microscopic extended Bose-Hubbard model with p-orbital degrees of freedom on a square lattice \cite{Liu_TSOC,Isacsson_pband}
\bea
&&\textstyle H = H_t +H_\text{onsite} , \nonumber \\
&&\textstyle H_t =
\sum_{\tbf{r}} {
   -t \left[a_{x} ^\dag (\tbf{r}) a_{x} (\tbf{r} +\hat{x})
    +a_{y} ^\dag (\tbf{r}) a_{y} (\tbf{r} +\hat{y})+ h.c. \right] }
\nonumber \\
 &&\textstyle - t_{\perp} \left[a_{x} ^\dag (\tbf{r}) a_{x} (\tbf{r} +\hat{y})
    +a_{y} ^\dag (\tbf{r}) a_{y} (\tbf{r} +\hat{x}) +h.c.\right] , \\
&&\textstyle H_\text{onsite} = \sum_{\tbf{r}}
 \frac{U}{2} ((n (\tbf{r}) (n (\tbf{r}) -\frac{2}{3}) -
\frac{1}{3} L_z (\tbf{r})^2) -\mu n.
%(\tbf{r}).
\eea
{Here, $a_{x}^\dag (\tbf{r})$ and
$a_{y} ^\dag (\tbf{r})$
are bosonic creation operators of p$_x$ and p$_y$ orbitals at $\tbf{r}$.}  The discrete variable $\tbf{r}$ labels the sites of a square lattice. The lattice constant $a$ is set to be $1$. $t$ ($t_\perp$) is the longitudinal (transverse) hopping between nearest neighbor sites, $U$ the on-site repulsion and the local angular momentum operator $L_z =-ia_y ^\dag a_x+i a_x ^\dag a_y$. The average occupation number $n$ of bosons per site is fixed by the chemical potential $\mu$.

Because $t<0$ and $t_\perp >0$, both of p$_x$ band and p$_y$ band show minima at finite momenta. It is inconvenient to take the long wavelength limit of the original lattice boson fields. To overcome this inconvenience we introduce the following staggered transformation
\be
\left[ \begin{array} {c}
\psi_{x} ^\dag (\tbf{r}) \\
\psi_y ^\dag (\tbf{r})
\end{array} \right]
=
\left[ \begin{array} {c}
(-1)^x a_{x} ^\dag  \\
(-1)^y a_{y} ^\dag
\end{array} \right],
\left[ \begin{array} {c}
\psi_\uparrow ^\dag(\tbf{r}) \\
\psi_\downarrow ^\dag(\tbf{r})
\end{array} \right]
=
\left[ \begin{array} {c}
\psi_{x} ^\dag +i \psi_{y} ^\dag\\
\psi_{x} ^\dag -i \psi_{y} ^\dag
\end{array} \right].
\label{eq:staggertransform}
\ee
$\psi_\uparrow ^\dag (\tbf{r}) $ and  $\psi_\downarrow ^\dag (\tbf{r})$ are lattice field operators for pseudospin
$|\uparrow(\tbf{r}) \rangle = (-)^x |p_x \rangle + i(-) ^y |p_y \rangle$ and $|\downarrow(\tbf{r}) \rangle = (-)^x |p_x\rangle - i (-)^y |p_y\rangle$ components,
{where $|p_x\rangle$ and $|p_y\rangle$ are local p$_x$ and p$_y$ orbital states.}

In this pseudospin representation, the Hamiltonian reads
\bea
\textstyle  H &=& \sum_{\tbf{r}, \tbf{r}'} T_{\sigma \sigma'} (\tbf{r} -\tbf{r}')
\psi_\sigma ^\dag (\tbf{r}) \psi_{\sigma'} (\tbf{r}') \nonumber \\
\textstyle &+& \sum_{\tbf{r}} \frac{U}{2} (n (\tbf{r}) ^2 -\frac{2}{3} n(\tbf{r}) -
\frac{1}{3} L_z (\tbf{r})^2)
-\mu n(\tbf{r}),
\label{Hor}
\eea
with
\bea
\textstyle &&T(\hat{x}) = \left[ \begin{array} {cc}
\frac{t- t_{\perp}}{2} & \frac{t +t_{\perp}}{2} \\
\frac{t+t_{\perp}}{2} & \frac{t -t_{\perp}}{2}
\end{array} \right], \nonumber \\
\textstyle &&T(\hat{y}) = \left[ \begin{array} {cc}
\frac{t- t_{\perp}}{2} & -\frac{t +t_{\perp}}{2} \\
-\frac{t +t_{\perp}}{2} & \frac{t-t_{\perp}}{2}
\end{array} \right],
\eea
{where $\sigma = \uparrow, \downarrow$, $n(\tbf{r}) = \psi_\uparrow ^\dag (\tbf{r}) \psi_\uparrow (\tbf{r})
+\psi_\downarrow ^\dag (\tbf{r}) \psi_\downarrow (\tbf{r})$, and
$L_z (\tbf{r}) = (-1)^{x+y} [\psi_\uparrow ^\dag (\tbf{r}) \psi_\uparrow (\tbf{r})
-\psi_\downarrow ^\dag (\tbf{r}) \psi_\downarrow(\tbf{r})]$. }
The fourier transform of $T_{\sigma \sigma'}$ gives
\bea
\epsilon(\tbf{k}) = \left[ \begin{array} {cc}
\epsilon_{\uparrow \uparrow} (\tbf{k}) %-(t - t_{\perp})(\cos(k_x) +\cos(k_y))
&  \epsilon_{\uparrow \downarrow} (\tbf{k})  \\ %-(t +t_{\perp})(\cos(k_x) -\cos(k_y)) \\
\epsilon_{\downarrow \uparrow}(\tbf{k})  %-(t +t_{\perp})(\cos(k_x) -\cos(k_y) )
&\epsilon_{\downarrow \downarrow} (\tbf{k}) % -(t -t_{\perp}) (\cos(k_x) +\cos(k_y) )
\end{array}\right],
\eea
where $\epsilon_{\uparrow \uparrow} (\tbf{k}) = \epsilon_{\downarrow \downarrow} (\tbf{k})
= (t - t_{\perp})(\cos(k_x) +\cos(k_y)) $ and
$\epsilon_{\uparrow \downarrow} (\tbf{k})  =\epsilon_{\downarrow \uparrow}(\tbf{k})
= (t +t_{\perp})(\cos(k_x) -\cos(k_y))  $.

It can be verified that the band structure given by $\epsilon(\tbf{k})$ shows a minimum at zero momentum. Thus we have obtained a theory which is convenient for us to take the continuum limit.
{The internal symmetry group of the Hamiltonian is U(1) $\times$ T, where T denotes the time reversal symmetry
($\psi_{\uparrow(\downarrow)} (\tbf{x},t) \to \psi_{\downarrow(\uparrow)} (\tbf{x}, -t)$) and U(1) denotes the global phase rotation symmetry ($\psi_\sigma \to e^{i\theta} \psi_\sigma$).} %[Do we use T or Z$_2$ for time reversal ? It seems
%T is more widely used. ]}
\begin{figure}
\centering
\includegraphics[angle=0,width=1.1\linewidth]{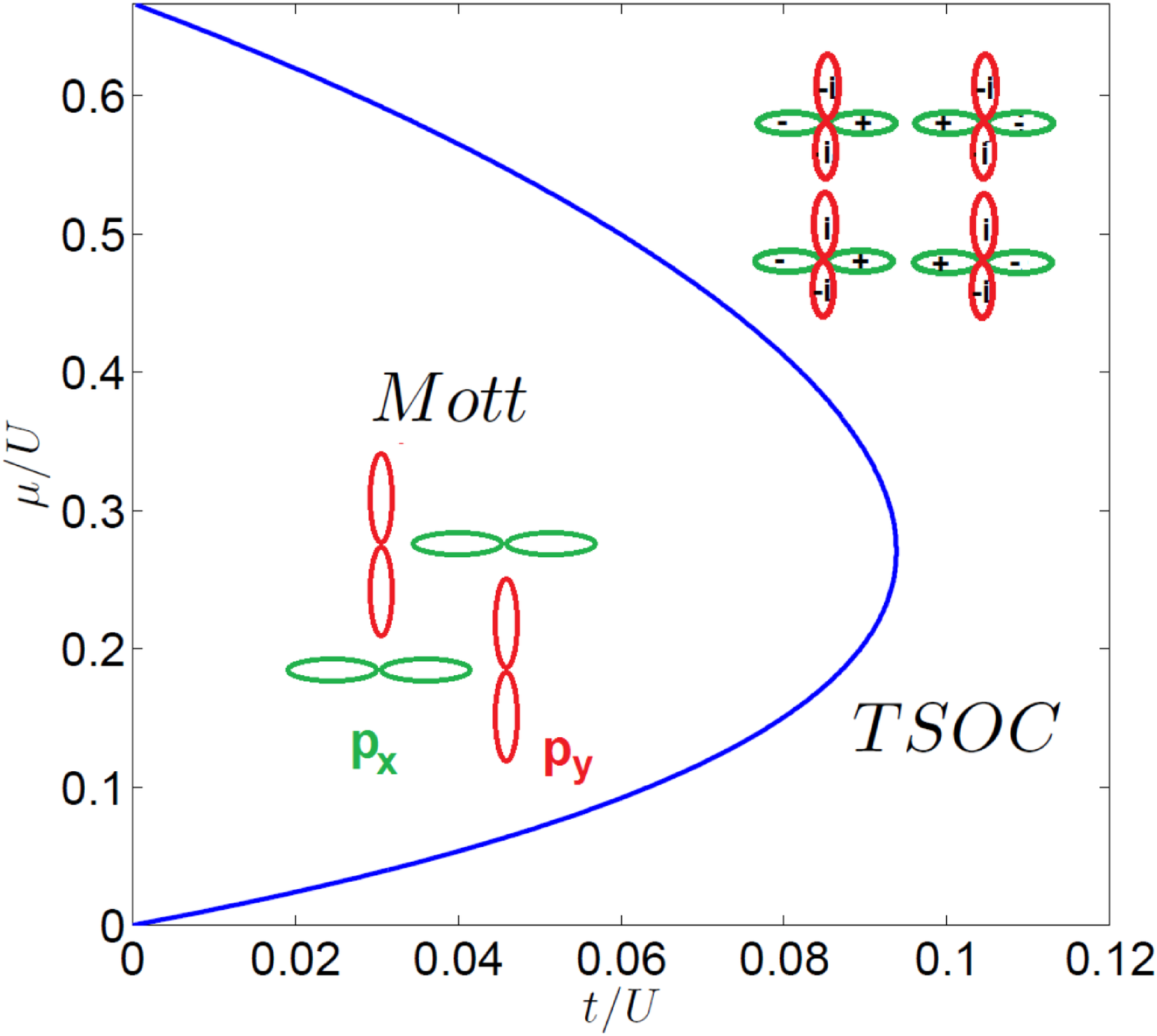}
\caption{(Color online). The phase diagram determined by
{ the effective action method.}
%{\bf mean field theory [mean field action derived from the effective action].}
The filling factor $\nu$ of the Mott regime shown is $1$.
$t_\perp$ is set to be $0.1 t$.
TSOC means the transversely staggered orbital current superfluidity\cite{Liu_TSOC}.
{The alternating p$_x$-p$_y$ pattern
shown in the Mott regime is the pattern of the Mott insulator with filling of $\nu=1$ in the p-bands. The staggered
p$_x\pm i$p$_y$ pattern in the TSOC regime illustrates the orbital current order in the TSOC phase. }
}
\label{fig:phasediag}
\end{figure}

{In the strong coupling limit, the system is in a Mott insulating phase for commensurate filling (filling factor
$\nu$ is an integer), where the filling is defined as the occupation number of bosons loaded on p orbits per site.
 %and U(1) symmetry cannot be broken due to the single particle gap induced by interaction.
 For filling factor $\nu$ larger than $1$, the interaction term favors local orbital current states because of the $(-L_z ^2)$ term in Hamiltonian~(Eq.~\ref{Hor}); these vortex-like states form a vortex-antivortex pattern due to super-exchange~\cite{Isacsson_pband,Collin_pband}.}
{For filling factor $\nu=1$,
the local vortex-like states are no longer favorable because the interaction term does not contribute on single particle states. Mathematically, the operator $L_z ^2$ term is equal to identity, when acting on the single particle states, and thus does not favor vortex-like states.
%the interaction term is locally SU(2) invariant (no longer favors local vortex states)
%because the operator $L_z ^2 $ is equal to identity on single particle states.
The Mott phase has an antiferro-orbital order, i.e., an alternating p$_x$-p$_y$ pattern (FIG.~\ref{fig:phasediag}), which breaks lattice translation symmetry~\cite{Isacsson_pband}.}
%But the translation by two times the original lattice constant is still a symmetry of this Mott state, thus we do not consider the effect of this broken lattice translation symmetry because
We are interested in the long wavelength modes within this phase.

{In the weak coupling limit, the system is in superfluid phase and the dispersion (obtained by diagonalizing $\epsilon(\tbf{k})$) shows minima at zero momentum.} %and the system is in superfluid phase.}
The two minimal single particle states carry lattice momentum $\tbf{k} =0$ and pseudospin $\sigma =  \uparrow, \downarrow$,  and they are related by time reversal (T) transformation. Due to the $(-L_z ^2)$ term in Hamiltonian, {the T symmetry is spontaneously broken in the ground state, i.e., either
$\langle \psi_\uparrow \rangle$ or $\langle \psi_\downarrow \rangle$ is finite. It is clear from
Eq.~(\ref{eq:staggertransform}) that
the original particles form a staggered p$_x\pm$ip$_y$ pattern (FIG.~\ref{fig:phasediag}) in this superfluid phase,
which is named TSOC~\cite{Liu_TSOC}.}
%The superfluid phase of this model we are exploring is a TSOC (FIG.~\ref{fig:phasediag})
%superfluid phase\cite{Liu_TSOC}.
Thus {going} from the Mott
insulator phase with filling $\nu =1$ to the TSOC superfluid phase, the U(1) $\times$ T symmetry is spontaneously broken. The phase transition (FIG.~\ref{fig:phasediag}) is confirmed by Gutzwiller mean field calculations~\cite{Isacsson_pband}. However, the momentum distribution and the {correlation functions} in the Mott phase are out of reach within Gutzwiller mean field calculation. {Motivated by this, we develop a theory valid in the strong coupling regime.}  With this theory we calculate the single particle spectrum for both of the  Mott phase and the TSOC superfluid phase and discuss how the Mott gap closes at the critical point and how the phase coherence peak develops in the Mott phase, and we explain the isotropy of the sound velocity of the TSOC superfluid phase in the strong coupling regime.

\section{Effective Action }
To capture the main feature of p-band Mott insulator and the TSOC superfluid in the strong coupling regime, {we aim at a theory capable of incorporating the local Mott gap, which is the leading effect of the Hubbard interaction, in a non-perturbative manner.} To do this, we follow the procedure in Ref.~\cite{Sengupta_hubbard,Hoffmann_hubbard,Freericks_hubbard}.
We first write the partition function $Z$ as a functional integral over complex fields $\psi_\sigma$ with the action $S[\psi^*, \psi] = \int_0 ^\beta d\tau \sum_{\sigma,\tbf{r}} { \{\psi_\sigma(\tbf{r}) \partial_\tau \psi_\sigma(\tbf{r}) +H[\psi^*, \psi]\} }$. We introduce an auxiliary field $\phi_\sigma$ to decouple the inter-site hopping term by means of a Hubbard-Stratonovich transformation and obtain
\bea
\textstyle Z &=& \int D[\psi_\sigma^*,\psi_\sigma,\phi_\sigma^*, \phi_\sigma]
e^{  \phi_\sigma ^* T^{-1} _{\sigma \sigma'} \phi_{\sigma'}
+[(\phi|\psi) +c.c.] -S_0 [\psi^*, \psi]  }  \nonumber \\
\textstyle &=& Z_0 \int D [\phi_\sigma^*, \phi_\sigma] e^{     \phi_\sigma ^* T^{-1} _{\sigma \sigma'} \phi_{\sigma'}   }
\langle \exp[(\phi|\psi) +c.c.] \rangle_0  \nonumber \\
\textstyle &\equiv& Z_0 \int D [\phi_\sigma^*, \phi_\sigma] \exp \left(    \phi_\sigma ^* T^{-1} _{\sigma \sigma'} \phi_{\sigma' }
+W[\phi_\sigma^*, \phi_\sigma] \right),
\label{eq:Zphi}
\eea
where the shorthand notation $(\phi|\psi) = \sum_{\sigma, \tbf{r}} \int_0 ^\beta d\tau
\phi_{\sigma}^*(\tbf{r}) \psi_\sigma (\tbf{r})$ and $T^{-1}$ denotes the inverse of the hopping matrix.
$S_0$ and $Z_0$ are the action and partition function in the local limit ($t, t_\perp =0$). $\langle \ldots \rangle_0$
means averaging over the local action $S_0 [\psi_\sigma ^* , \psi_\sigma]$. The introduced
generating function $W[\phi_\sigma ^*, \phi_\sigma] =\ln \langle \exp \left(  (\phi|\psi) +c.c.\right) \rangle_0$.
The local action $S_0$, which is equivalent to the original action without tunneling term, is invariant under a $U(1) \times U(1)$ transformation
$\psi_\sigma \to e^{i \theta_\sigma} \psi_\sigma$. 
{%\color{red} 
By definition, $W[\phi_\sigma ^*, \phi_\sigma]$ is independent of hopping ($t$ and $t_\perp$).} 
Power expansion of $W[\phi_\sigma ^*, \phi_\sigma]$ respecting this symmetry yields
%$\phi_{a _2}$
\bea
\textstyle &&W = \int d\tau_1 d \tau_2 \sum_{\tbf{r}} \text{G} _\sigma (\tbf{r}, \tau_1 -\tau_2)
{\phi}^*_\sigma (\tbf{r},  \tau_1) \phi_{\sigma} (\tbf{r}, \tau_2)  \nonumber \\
\textstyle &&+ \frac{1}{2!} \int \prod_{\alpha =1} ^{4} d\tau_\alpha \sum_{\tbf{r}} \nonumber \\
\textstyle && ~~~~~~{\chi}_{\sigma_1 \sigma_2}
(\tbf{r}, 1234) {\phi}^*_{\sigma_1} ( 1) \phi_{\sigma_1} ( 2)
{\phi}^*_{\sigma_2} (  3) \phi_{\sigma_2} ( 4) \nonumber \\
\textstyle &&+O(\phi^6) ,
\label{eq:W}
\eea
where the indices $1,2,3$, and $4$ indicate time $\tau_1, \cdots, \tau_4$. And because of T (time reversal) symmetry,
$\text{G} _\uparrow = \text{G} _\downarrow$,
 and $\chi _{\uparrow \uparrow} =\chi _{\downarrow \downarrow}$.
 {
 In the transformed theory, the quadratic term of $T^{-1}$ in Eq.~(\ref{eq:Zphi}) is dominant in the strong coupling limit, $t/U\rightarrow 0$. We thus truncate the power expansion to quartic order. Now that the fluctuations of $\phi$ fields are controlled by $T^{-1}$ (Eq.~(\ref{eq:Zphi})), perturbative renormalization group (RG) analysis finds higher order terms are irrelevant.}
From the definition, the coefficients are readily obtained as follows
\bea
&&\text{G} _\sigma (\tbf{r}, \tau_1 -\tau_2) = \langle \psi_\sigma (\tbf{r},\tau_1) \psi^*_\sigma (\tbf{r}, \tau_2) \rangle^c _0 , \nonumber \\
&&\chi _{\sigma_1 \sigma_2} (\tbf{r};1234) =
\langle \psi_{\sigma_1} ( 1) \psi_{\sigma_1} ^* (2)  \psi_{\sigma_2 } (3) \psi_{\sigma_2} ^* (4) \rangle^c _0.
\label{cors0}
\eea
The effective action for $\phi$ fields is  $S_{\text{eff}} [\phi_\sigma^*, \phi_\sigma] = \int d\tau\sum_{\tbf{r}, \tbf{r}'} -\phi^* _\sigma (\tbf{r}) T^{-1}_{\sigma \sigma'} (\tbf{r}-\tbf{r}')
\phi_{\sigma'} (\tbf{r}') -W[\phi_\sigma ^*, \phi_\sigma]$,
which can be used as a starting point to study the instability of Mott phase with respect to superfluidity by treating $\phi$ fields as superfluid order parameters~\cite{Oosten_hubbard}.
However it is inconvenient to calculate the excitation spectrum and the momentum distribution from this action~\cite{Sengupta_hubbard}. 
{%\color{red} 
Also, the theory $S_{\text{eff}} [\phi_\sigma^*, \phi_\sigma] $ does not provide a clear picture of quasi particles.} The above difficulties can be overcome by performing a second Hubbard-Stratonovich transform following the method of Ref.~\cite{Sengupta_hubbard},
\be
Z =\int D[ \varphi ^* \varphi  \phi^* \phi ] e^{ -\varphi_\sigma ^* T_{\sigma \sigma'} \varphi_{\sigma'}
-[(\varphi|\phi)+c.c.] +W[\phi_\sigma^*,\phi_{\sigma}]
 }.
\ee
{Integrating out $\phi$ fields gives} the effective action $S_{\text{eff}} [\varphi_\sigma ^*, \varphi_\sigma] $
\bea
&&S_{\text{eff}} [\varphi_\sigma^*, \varphi_\sigma] = -\tilde{W}[\varphi^* _\sigma, \varphi_\sigma] +\int d\tau
\sum_{<\tbf{r}_1, \tbf{r}_2>}
\mathcal{L}_0 ,
\eea
with
\bea
&&\mathcal{L}_0 =\varphi_{\sigma_1} ^* (\tbf{r}_1, \tau)
T_{\sigma_1 \sigma_2} (\tbf{r}_1-\tbf{r}_2)
\varphi_{\sigma_2} (\tbf{r}_2, \tau) , \nonumber \\
&&\tilde{W} =\ln \langle \exp \left( -[(\varphi| \phi) +c.c. ]\right) \rangle_W,
\eea
where $\langle \ldots \rangle_W = \frac{\int D(\phi^*_\sigma, \phi_\sigma) (\ldots)
\exp({W[\phi^*_\sigma, \phi_\sigma]})}
{\int D(\phi^*_\sigma, \phi_\sigma  \exp({W[\phi^*_\sigma, \phi_\sigma   ]})}$. 
{%\color{red}
Since the functional $W[\phi_\sigma ^*, \phi_\sigma]$ is independent of hopping, $\tilde{W}[\varphi_\sigma ^*, \varphi_\sigma]$ is also independent of hopping by definition.} 

{Now the task is to calculate the functional $\tilde{W}$. The essence is to evaluate the expectation value of the exponential operator in a system described by the effective action $W$.
Note that the action $W$ (Eq.~(\ref{eq:W})) is a power expansion of the small parameter $1/U$ (strong coupling), so the perturbative field theoretical method is valid and powerful to compute the expectation value.} In this manner, we perform the power expansion of $\tilde{W}[\varphi^* _\sigma, \varphi_\sigma]$ respecting the $U(1) \times U(1) $  symmetry
and obtain the effective theory for $\varphi_\sigma$ as follows
\bea
&&S_{\text{eff}} [\varphi_\sigma^*, \varphi_\sigma] = \nonumber \\
&& \int d\tau_1 d\tau_2 \sum_{\tbf{r}}
\varphi_{\sigma} ^* (\tbf{r},\tau_1) \text{G}^{-1}_{\sigma} (\tau_1 -\tau_2)
	\varphi_{\sigma} (\tbf{r},\tau_2) \nonumber \\
&& ~~+\int d\tau
\sum_{<\tbf{r}_1, \tbf{r}_2>}  \varphi_{\sigma_1} ^* (\tbf{r}_1, \tau)
T_{\sigma_1 \sigma_2} (\tbf{r}_1-\tbf{r}_2)
\varphi_{\sigma_2} (\tbf{r}_2, \tau) \nonumber \\
&&~~+\frac{1}{2} g_{\sigma_1 \sigma_2} \int d\tau \sum_{\tbf{r}} { |\varphi_{\sigma_1} (\tbf{r},\tau)|^2 |\varphi_{\sigma_2}(\tbf{r}, \tau)|^2}.
\label{Seff}
\eea
Here we have truncated the power expansion to quartic order.
The vertex term can have a finite range behavior in $\tau$ space in principle, but we can take the static limit because the $\tau$ dependent corrections would be irrelevant in the RG sense.
{
The field $\psi$ describes the bare (original) bosons with onsite interaction $U$, while
the field $\varphi$ describes quasi particle excitations which are greatly suppressed by the
energy gap (of the order of $U$) in the strong coupling regime (this physical understanding comes from Eq.~(\ref{eq:gvarphi})). 
%The bare particles of $\psi$ are simply the original bosons which interact strongly for large
%$U$; while the bare particles of $\varphi$ defined by the bare two point correlator
%of $\varphi$ (Eq.~\ref{eq:gvarphi}) are particle/hole excitations on top of the Mott state.
%We can call these bare
%particles of $\varphi$ ---quasi particles.
%These quasi particles are dilute in the strong coupling
%regime and they play the key role in the Mott-superfluid transition.
Thus the quasi particles are dilute, and we expect the three body scattering process is negligible,
which further guarantees the validity of the truncation performed in Eq.~(\ref{Seff}). }
{%\color{red}
Here, we want to emphasize the quasi particles hop around through $T_{\sigma \sigma'}$ (Eq.~(\ref{Seff})). 
The quasi particles are mobile instead of localized. The dispersion of the quasi particles is discussed in the next section.} 

%Thus we will not distinguish $\varphi$ fields or $\psi$ fields and write the effective action
%$S_\text{eff}  [\varphi_\sigma^*, \varphi_\sigma]$ by
%$S_\text{eff}  [\psi_\sigma^*, \psi_\sigma]$.
{%\color{red} 
Despite the difference of bare correlators of $\psi$ and $\varphi$,
it is proved that the correlators defined by the original action $S[ \psi^*, \psi]$
are exactly equal to that defined by the infinite series of power expansion of this effective
action $S_\text{eff}  [\varphi_\sigma^*, \varphi_\sigma]$~(Appendix B). %~\cite{Sengupta_hubbard}.
%It is not mysterious that the double transformed theory has even more direct relations to the original theory. 
%It is established in field theory that the effective action generating 1PI correlation functions, 
%instead of the generating functional (generating the connected correlation functions) itself, has a more direct relation to the 
%original field theory. %The typical approach of calculating the effective action in field theory involves a Legendre transform of 
%the generating functional.  
%the exact connected correlators defined by this effective action
%$S_\text{eff}  [\varphi_\sigma^*, \varphi_\sigma]$ are the same as the
%correlators defined by the original action $S[ \psi^*, \psi]$ if we do not truncate
%the power expansions~\cite{Sengupta_hubbard}.
With valid truncations, the connected correlators of $\varphi$ reproduce the connected correlators
of $\psi$ approximately.
}
In principle, one can follow the above procedure and get $\text{G}^{-1} _{\sigma_1} (\tau_1 -\tau_2)$
and $g_{\sigma_1 \sigma_2}$. It is straightforward to calculate these coefficients for the one component Bose-Hubbard
Model. However it is inconvenient to proceed in this approach due to the complexity induced by the local degeneracy of
ground states of the action $S_0$. Since these coefficients are \emph{independent of hopping}, we decide to  calculate the coefficients
by identifying correlators of this effective action and the original correlators
(defined by the original action) in the local limit ($t, t_\perp =0$). The derivation and the results are summarized in Appendix A.

After the above manipulation, the leading effect of the Hubbard interaction, namely, generating a local energy gap in the quasi-particle spectrum, is included in the quadratic part of $S_{\text{eff}} [\varphi_\sigma^*, \varphi_\sigma]$, which makes the following Bogoliubov analysis valid in the strong coupling regime. 
{%\color{red} 
With the effective theory $S_{\text{eff}} [\varphi_\sigma^*, \varphi_\sigma]$, the superfluid phase in the strong coupling regime is described as a superfluid phase of weakly interacting quasi particles. We want to emphasize that this physical picture is lacking in the theory $S_{\text{eff}} [\phi_\sigma^*, \phi_\sigma]$. }

\section{Zero and finite temperature phase transitions}
In this paper, we focus on the lowest Mott lobe regime ($\nu =1$) for which time reversal T symmetry is not broken. We start our analysis from the effective action (Eq.~(\ref{Seff})).
We use the correlators of quasi particle fields $\varphi$, which
can be calculated within Bogoliubov theory, to approximate
the correlation functions and momentum distribution of the bare particles. 
(The reason is explained in Appendix B.) 
{%\color{red} 
The method, we apply here, yields results~\cite{Sengupta_hubbard} that agree with for example the RPA calculation~\cite{Freericks_hubbard}, when calculating the momentum distribution of the s-band Bose-Hubbard model. }
%Since the quartic term in this action is a perturbative term, we apply the Bogoliubov approach to study the Mott-superfluid phase transition and the quasi particle spectra in both phases.

First, let us determine the ground state which is homogeneous and static after the staggered transformation (Eq.~(\ref{eq:staggertransform})). The free energy functional is
%The mean field action derived from the effective action in Eq.~(\ref{Seff}) is given by
\be
\textstyle \frac{S}{N\beta} = (\mathcal{G}^{-1} _\sigma (0)+ \epsilon_{\sigma \sigma} (\tbf{k} =0)  ) n_{s, \sigma}
+\frac{1}{2}  g_{\sigma_1 \sigma_2}  n_{s, \sigma_1} n_{s, \sigma_2},
\ee
where $N$ is the number of lattice sites, $\mathcal{G}_\sigma (i \omega)$ is the fourier transform of $\text{G}_\sigma (\tau)$
and $n_{s, \sigma} = |\langle \varphi_{\sigma} \rangle |^2$ is the superfluid density of the pseudospin  $\sigma$
component.
By minimizing the free energy functional, we get
\be
n_{s, \sigma} =
\begin{cases}
-\frac{\mathcal{G}^{-1} _\sigma (0)+ \epsilon_{\sigma \sigma} (\tbf{k} =0) }{g_{\sigma \sigma}}	
	& \text{if $\mathcal{G}^{-1} _\sigma (0)+ \epsilon_{\sigma \sigma} (\tbf{k} =0) <0$;} \\
0 & \text{otherwise.}
\end{cases}
\label{sfdensity}
\ee
In the Mott regime, $n_{s, \sigma}$ vanishes. In the superfluid regime, only one component is finite, i.e., either $n_{s,\uparrow}$ or $n_{s,\downarrow}$ is finite and the other vanishes, because the off-diagonal part $g_{\uparrow \downarrow}$  is greater than the diagonal part $g_{\uparrow \uparrow}$  ($= g_{\downarrow \downarrow}$). Thus $U(1) \times T$ is spontaneously broken across the Mott-superfluid phase transition in this model for filling $\nu =1$. The phase boundary is shown in FIG.~\ref{fig:phasediag}. 
The phase diagram is consistent with Ref.~\cite{Isacsson_pband}. {The Mott regime determined by our approach is larger.}
{%\color{red}
 (When applied to calculate the phase boundary of s-band Bose-Hubbard model, the method adopted here yields results that agree with other mean field theories~\cite{Sengupta_hubbard}.)} In the following part, we assume the superfluid component is the `$\uparrow$' component.
{The superfluid density scales as
$n_{s, \uparrow} \sim |n-\nu|^\zeta$ away from the Mott tip when the mean particle
number per site, $n$, is not equal to $\nu$, whereas it scales as $n_{s,\uparrow} \sim (t-t_c)^{\zeta'}$ at the Mott tip where $n = \nu$. }

Next, we explore the fluctuations $\tilde{\varphi}_{\sigma} (\tbf{r}, \tau) = \varphi_{\sigma} (\tbf{r},\tau) - \sqrt{n_{s,\sigma}}$.
Expanding the action to the quadratic order of the fluctuation fields $\tilde{\varphi}_{\sigma}$, we get
\be
S[\tilde{\varphi}_\sigma ^*, \tilde{\varphi}_\sigma] = \frac{1}{2} \sum_{\tbf{k}, \omega} \Psi^\dag(\tbf{k}, i\omega) [h(\tbf{k}, i\omega)] \Psi(\tbf{k}, i\omega),
\ee
with
\be
\Psi^\dag (\tbf{k}, i\omega) = \left[ \tilde{\varphi} ^* _\uparrow (\tbf{k}, i\omega),
	\tilde{\varphi} _\uparrow (-\tbf{k}, -i\omega),
	\tilde{\varphi} ^* _\downarrow (\tbf{k}, i\omega),
	\tilde{\varphi} _\downarrow (-\tbf{k}, -i\omega) \right]
\ee
%\be
%\Psi^\dag (\tbf{k}, i\omega) = \left[ \tilde{\psi} ^* _\uparrow (\tbf{k}, i\omega),
%	\tilde{\psi} _\uparrow (-\tbf{k}, -i\omega),
%	\tilde{\psi} ^* _\downarrow (\tbf{k}, i\omega),
%	\tilde{\psi} _\downarrow (-\tbf{k}, -i\omega) \right]
%\ee
and
\bea
h_{11} &=& \epsilon_{\uparrow \uparrow} (\tbf{k}) + \mathcal{G}_\uparrow ^{-1} (i\omega)
	+2 g_{\uparrow \uparrow} n_s ~~,\nonumber \\
h_{12} & = & h_{21} =  g_{\uparrow \uparrow}  n_s~~, \nonumber \\
h_{13} & = & h_{31} = \epsilon_{\uparrow \downarrow} (\tbf{k}) ~~,\nonumber \\
h_{22} & =& \epsilon_{\uparrow \uparrow} (-\tbf{k}) + \mathcal{G}_\uparrow ^{-1} (-i\omega)
	+2 g_{\uparrow \uparrow} n_s ~~, \nonumber \\
h_{24} & = & h_{42} = \epsilon_{\downarrow \uparrow} (-\tbf{k}) ~~,\nonumber \\
h_{33} & = &  \epsilon_{\downarrow \downarrow} (\tbf{k}) + \mathcal{G}_\downarrow ^{-1} (i\omega)
	+g_{\uparrow \downarrow} n_s ~~,\nonumber \\
h_{44} & = & \epsilon_{\downarrow \downarrow} (-\tbf{k}) + \mathcal{G}_\downarrow ^{-1} (-i\omega)
	+ g_{\uparrow \downarrow} n_s ~~,
\label{hmat}
\eea
where $\tilde{\varphi}_\sigma (\tbf{k}, i\omega)$ is the fourier transform of $\tilde{\varphi}_\sigma(\tbf{r},\tau)$.
Only non-zero $[h]$ matrix elements are listed above. Using the quadratic action for fluctuations, we calculate the Bogoliubov spectrum.

\subsection{Mott phase}
In this part, we study the momentum distribution of p-band Mott insulator phase and we show that diverging peaks for bosons on p$_x$ and p$_y$ bands rise at finite momenta when the Mott gap closes. This offers new approaches of preparing coherent matter waves in experiments.

\begin{figure}
\centering
\includegraphics[angle=0,width=0.49\linewidth]{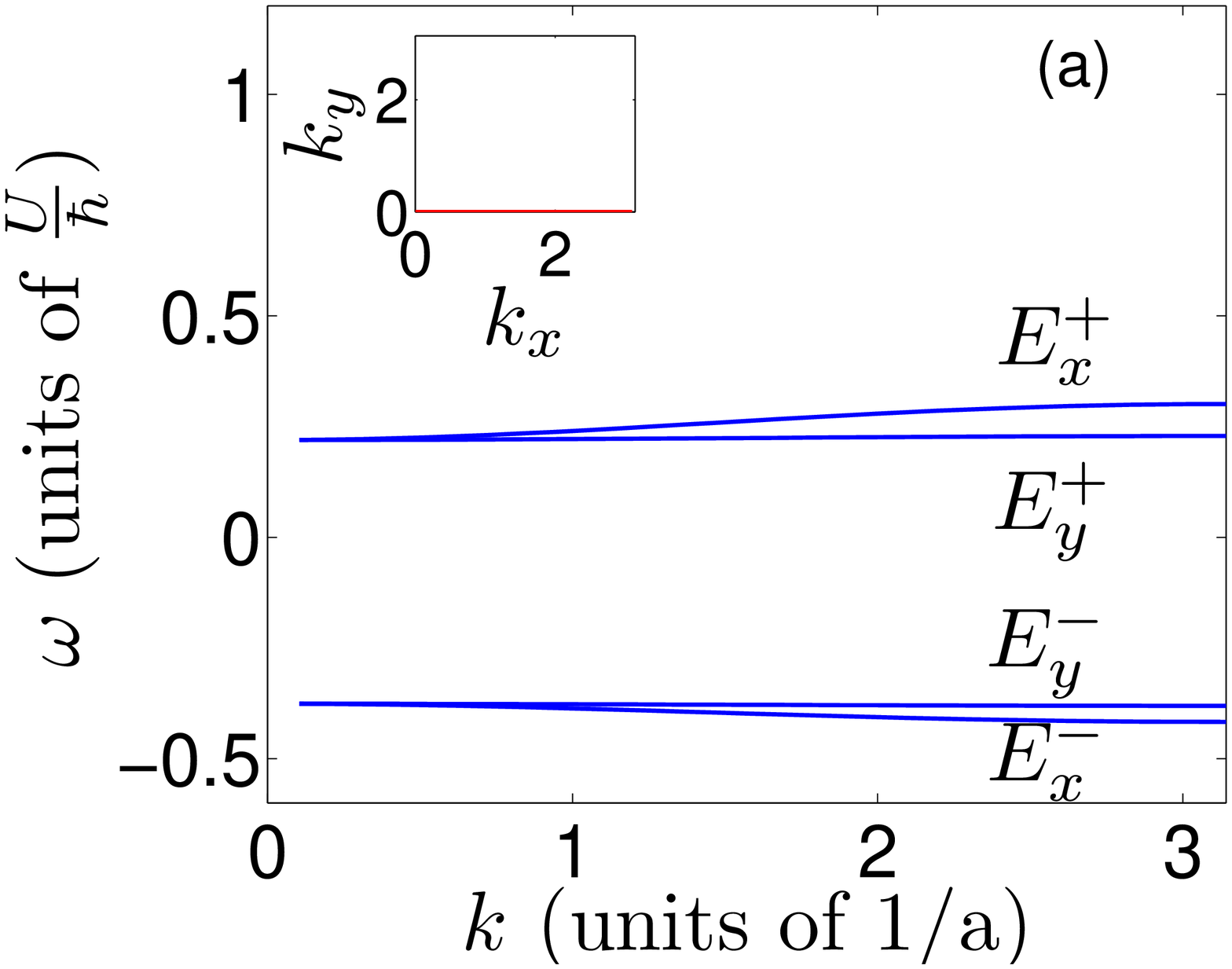}
%    \label{fig:spectramott1a}}
%\qquad
\includegraphics[angle=0,width=0.49\linewidth]{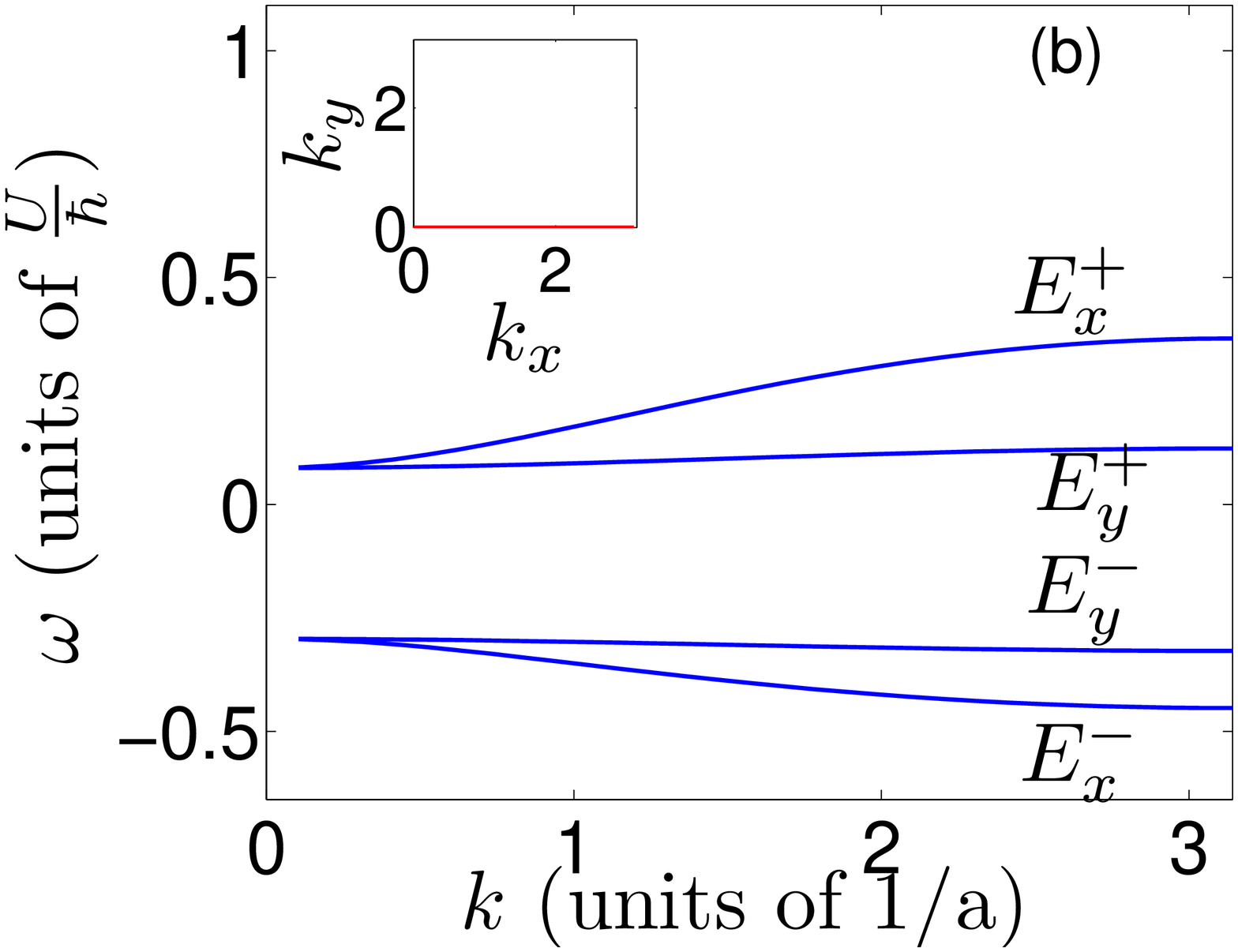}
%    \label{fig:spectramott1b}}
\caption{(Color online). (a) The single particle spectra along the $k_x$ axis
%$\tbf{k} = k \hat{x}$ direction
{(shown in the inset)}
deep in the Mott regime with parameters
$\mu/U =0.4, t/U =0.02, t_\perp  = 0.1t$.
(b) The single particle spectra along the $k_x$ axis
%$\tbf{k} = k\hat{x}$ direction
{(shown in the inset)}
in the Mott regime near the critical point with parameters
$\mu/U =0.4, t/U =0.065, t_\perp = 0.1t$. The Mott gap drops when increasing hopping and the gap of particle branches closes at the critical point for $\mu/U = 0.4$.
}
\label{fig:spectramott1}
\end{figure}

\begin{figure}
\includegraphics[angle=0,width=0.49\linewidth]{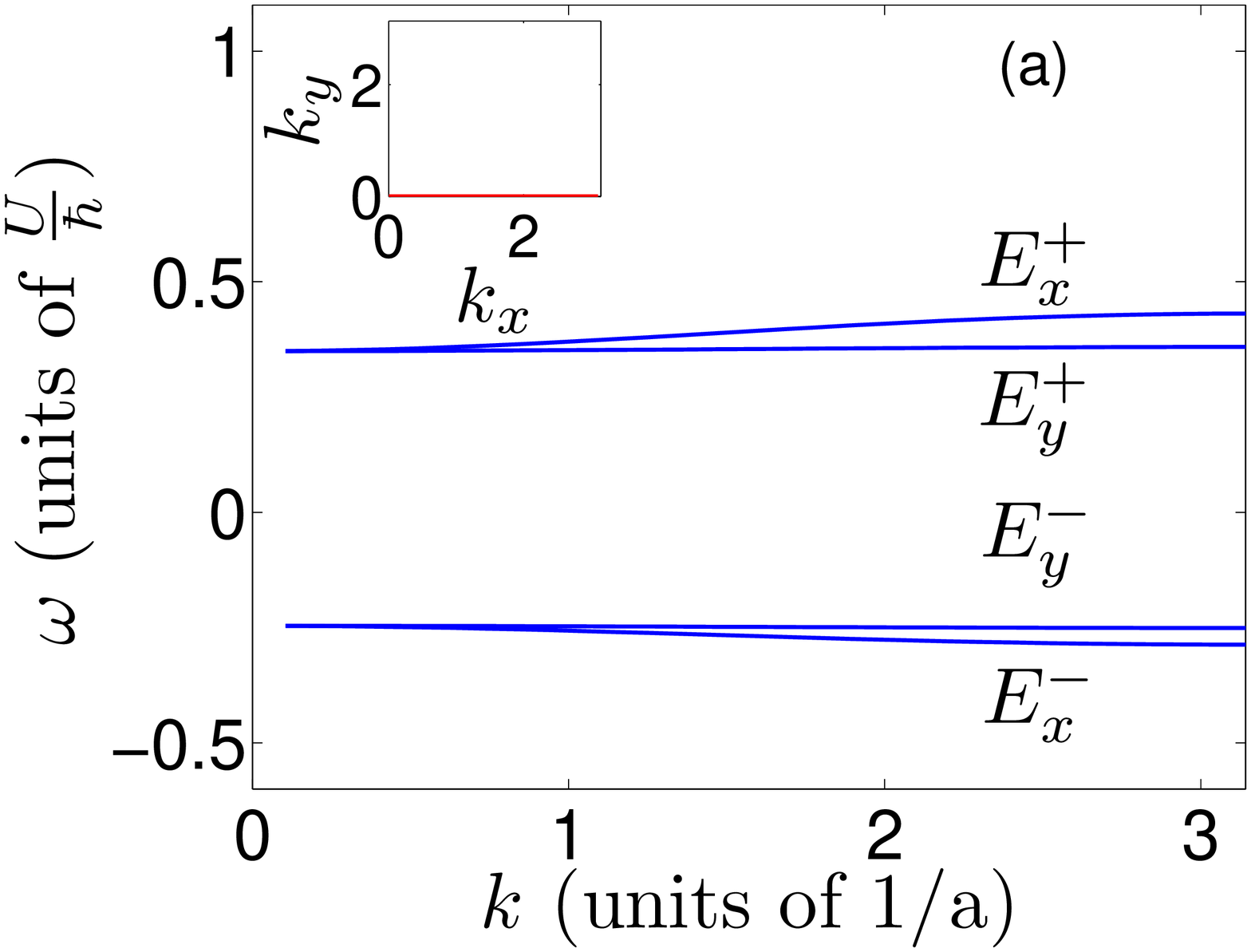}
\includegraphics[angle =0, width =0.49\linewidth]{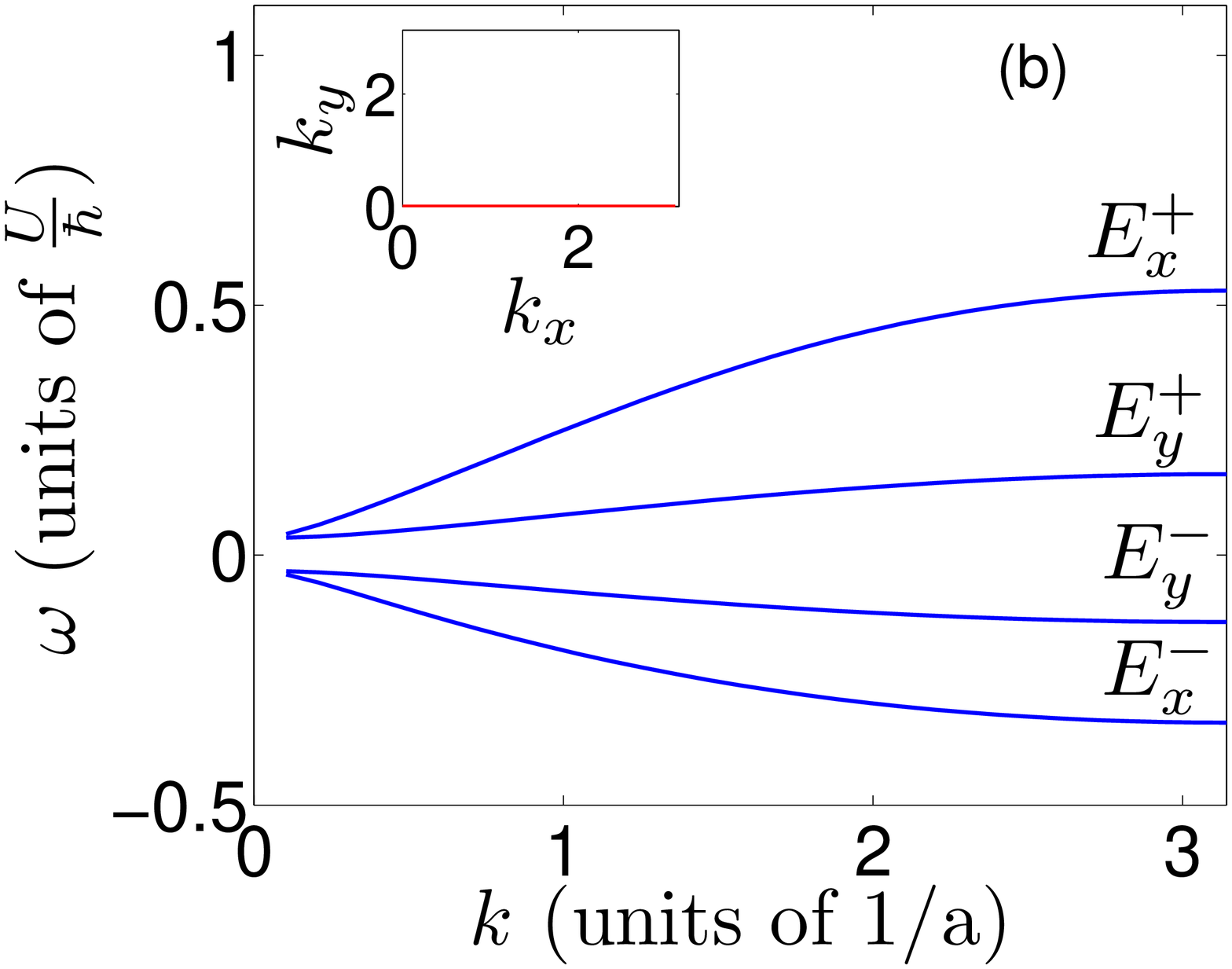}
\caption{(a) The single particle spectra along $k_x$ axis deep in the Mott regime with parameters
$\mu/U =0.27, t/U =0.02, t_\perp  = 0.1t$.
(b) The single particle spectra along the $k_x$ axis in the Mott regime near the critical point with parameters
$\mu/U =0.27, t/U =0.09, t_\perp = 0.1t$.  The Mott gap for both of the particle and hole branches closes at the critical point at the Mott tip regime.
}
\label{fig:spectramott2}
\end{figure}

In the Mott phase, superfluid density $n_s =0$. The Green function
$\mathscr{G}_{\sigma_1 \sigma_2} (\tbf{k}, i\omega) = \langle \psi_{\sigma_1} (\tbf{k}, i\omega )\psi_{\sigma_2}^*(\tbf{k}, i\omega) \rangle$ $= \langle \varphi_{\sigma_1} (\tbf{k}, i\omega )\varphi_{\sigma_2}^*(\tbf{k}, i\omega) \rangle$
is readily obtained as $\mathscr{G} ^{-1} _{\sigma_1 \sigma_2} (\tbf{k}, i\omega) =
\epsilon_{\sigma_1 \sigma_2} (\tbf{k} )+ \mathcal{G}_{\sigma_1}^{-1} (i\omega) \delta_{\sigma_1 \sigma_2}$.
Solving the equation $\det [\mathscr{G} ^{-1} (\tbf{k}, \omega) ] =0$, we get the single particle spectra shown in
FIG.~\ref{fig:spectramott1} and FIG.~\ref{fig:spectramott2}.
%The alternating $p_x$ $p_y$ pattern \cite{Isacsson_pband} of the Mott state provide a staggered background
%potential for single particle excitations, which will affect modes with $\tbf{k} =(\pm \pi, \pm \pi)$.
%This background has a more important effect on the hole excitations~\cite{note2}. Our result should be
%taken as the staggered background potential is coarse grained.
Deep in the Mott regime, all single particle excitations are fully gapped. It can be verified that the Green function $\mathscr{G}_{\sigma_1, \sigma_2} (\tbf{k}, i\omega)$ is diagonal in the \{$\psi_x (\tbf{k}, i\omega)$, $\psi_y (\tbf{k}, i\omega)$\} basis. Thus we can label the spectra by $E_x ^\pm$ and $E_y ^\pm$. The $E_x ^+ (E_x ^-) $ branch is the p$_x$ band of particle (hole) excitations; while $E_y ^+ (E_y ^- ) $ branch is the p$_y$ band of particle (hole) excitations. Upon the phase transition point, the gap at $\tbf{k}=0$ drops and approaches zero. Away from the Mott tip, two particle branches ($E_x ^+$ and $E_y ^+$) touch zero first (FIG.~\ref{fig:spectramott1}), which causes the instability of the Mott insulator phase and drives the phase transition from Mott insulator to TSOC superfluid. For a more physical case in which the density is fixed to be an integer, both of particle and hole branches close simultaneously when increasing hopping because of the particle-hole symmetry at the Mott tip (FIG.~\ref{fig:spectramott2}). For both cases the gap closes right at the phase transition point determined by minimizing the free energy functional in Eq.~(\ref{sfdensity}).

\begin{figure}
\subfigure[ ]{
    \includegraphics[angle=0,width=0.4\linewidth]{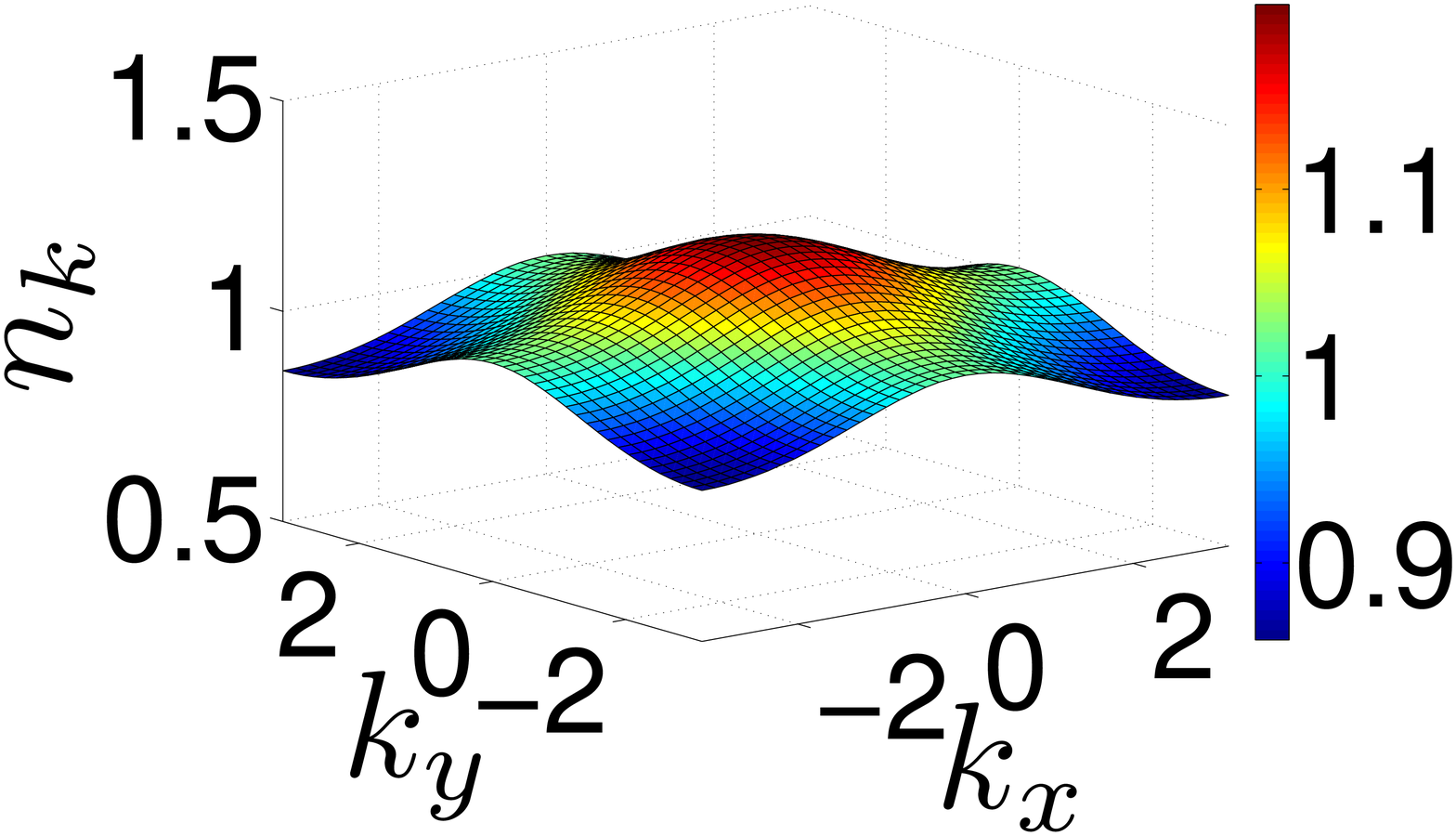}
    \label{fig:momdistmott1}
}
\subfigure[ ] {
    \includegraphics[angle =0, width =0.4\linewidth]{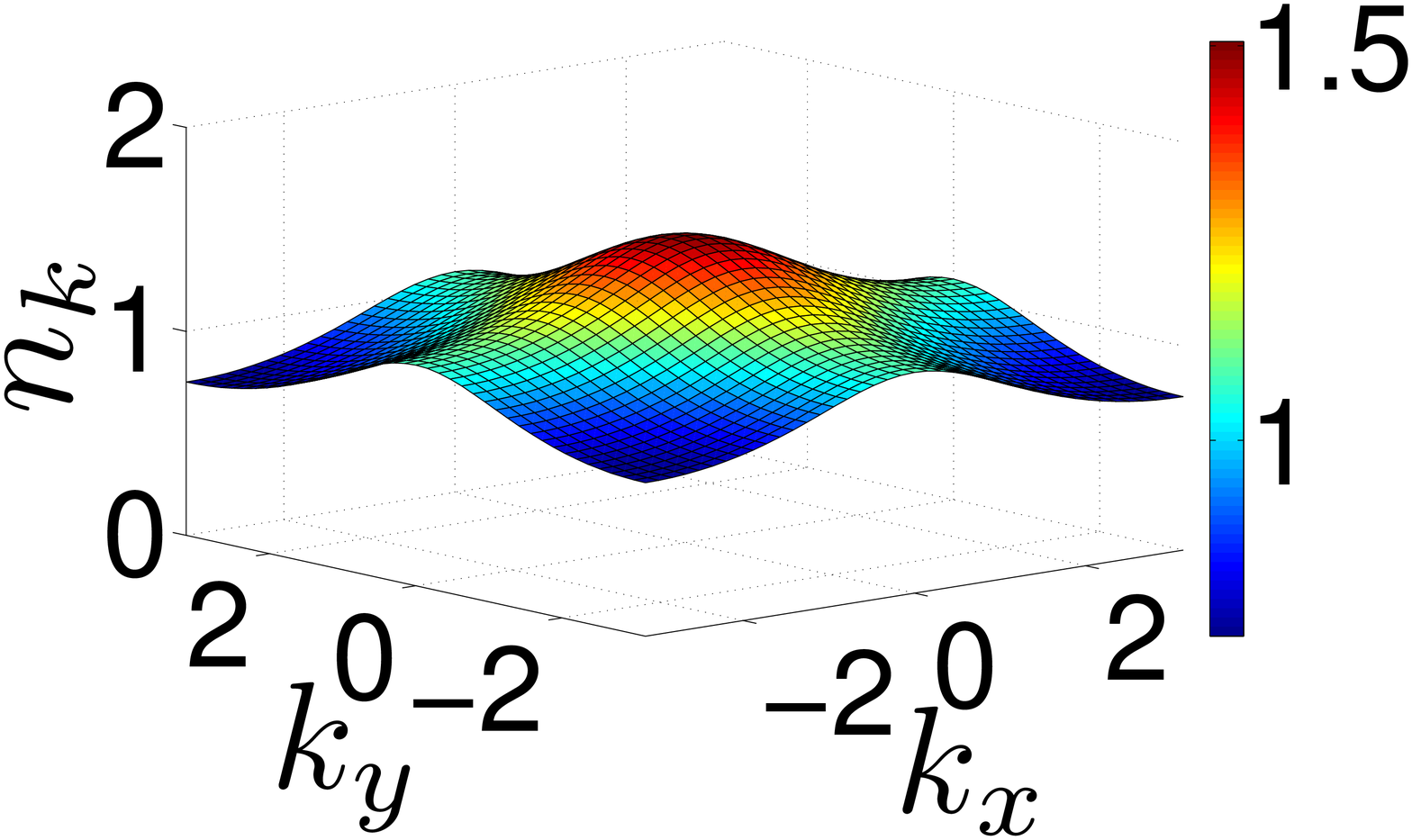}
    \label{fig:momdistmott11}
}
\subfigure[ ] {
    \includegraphics[angle =0, width =0.4\linewidth]{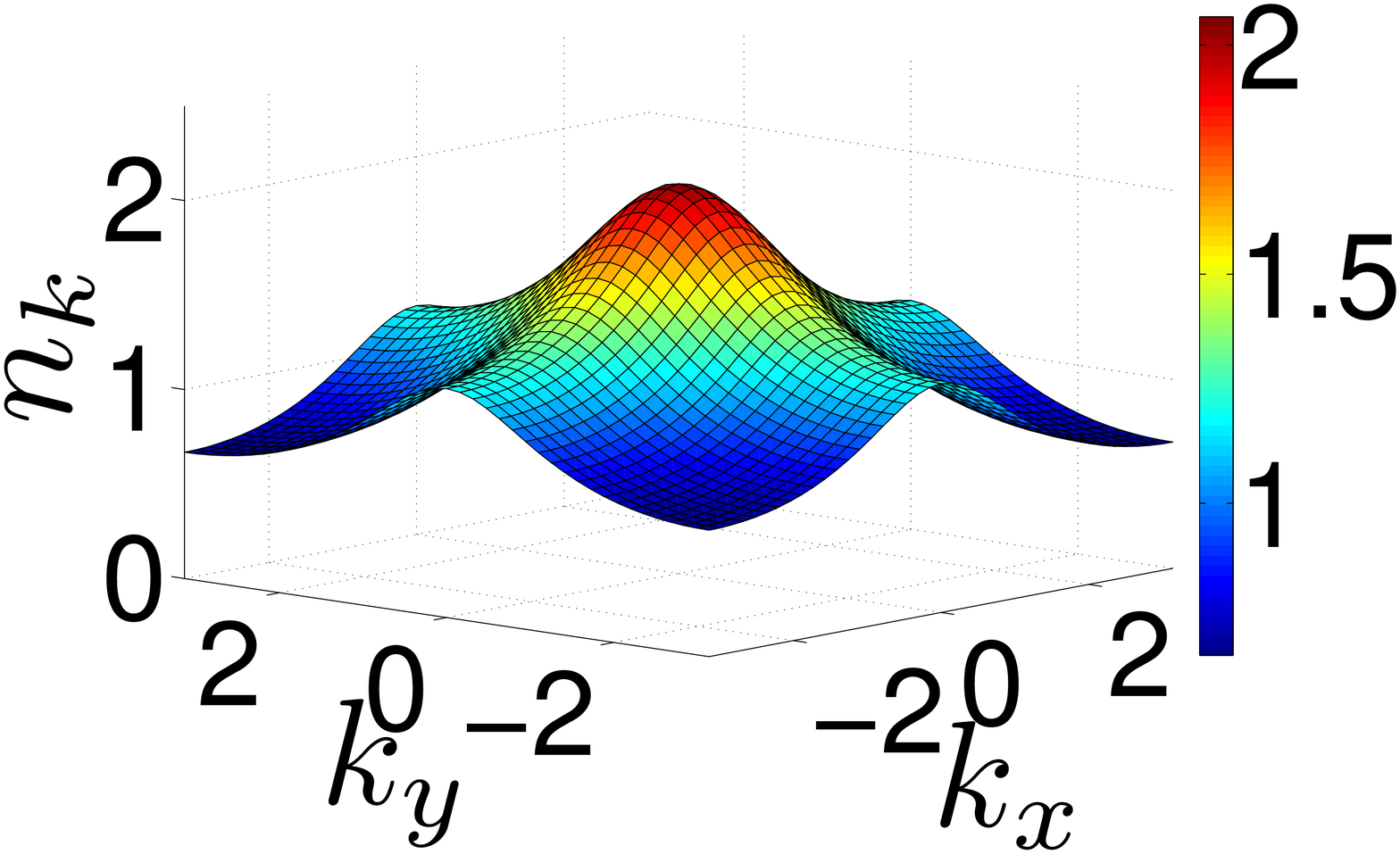}
    \label{fig:momdistmott21}
}
\subfigure[ ] {
    \includegraphics[angle =0, width =0.4\linewidth]{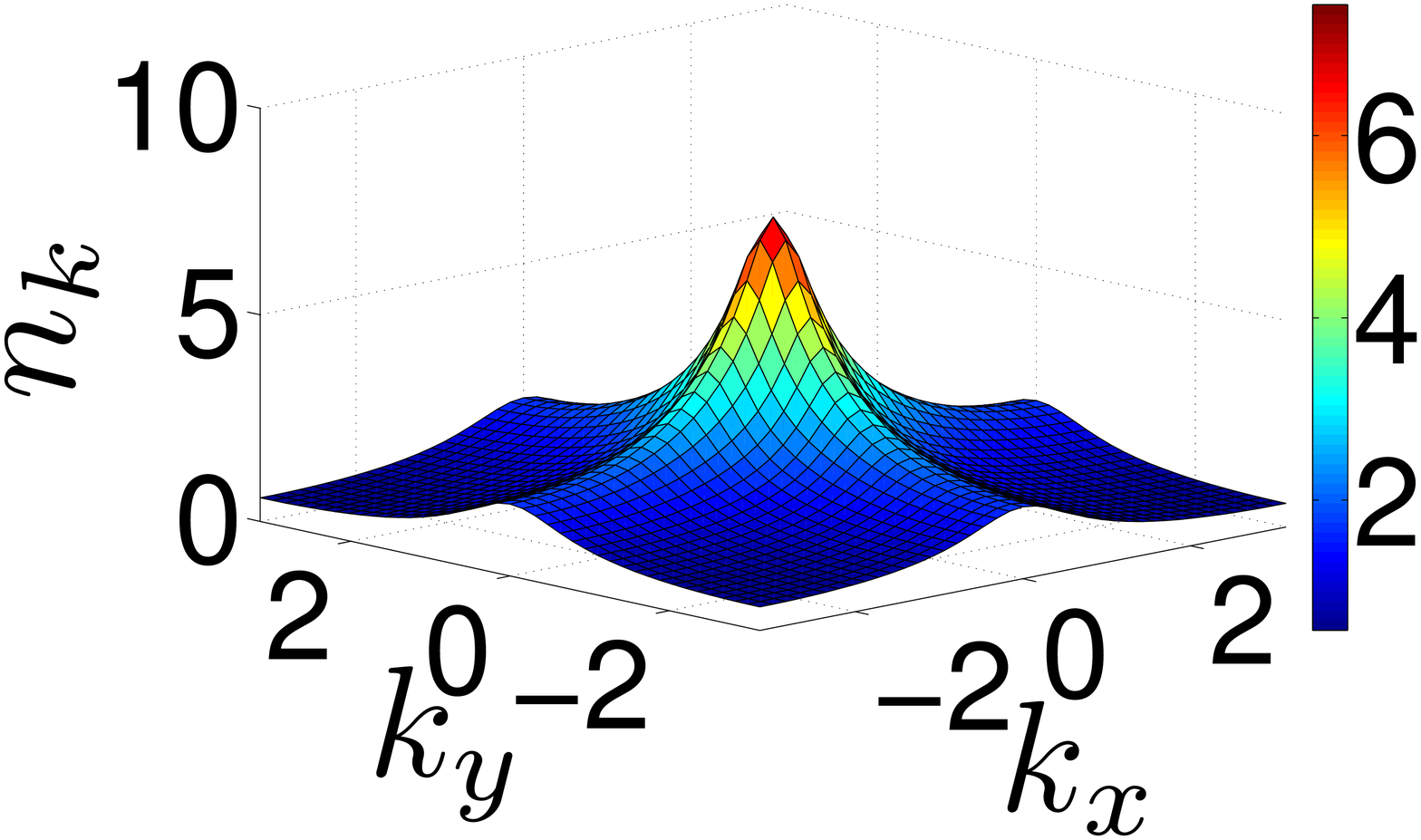}
    \label{fig:momdistmott2}
}
\caption{(Color online). The momentum distribution defined by Tr$\langle\psi_{\sigma_1}^* (\tbf{k})  \psi_{\sigma_2} (\tbf{k}) \rangle$ in the Mott regime with
parameters $\mu/U =0.3, t_\perp  = 0.1t$. For (a) through (d), the parallel hopping $t/U$ is 0.02, 0.04, 0.06 and 0.09 respectively. A coherent peak rises continuously when approaching the Mott-superfluid transition point from the Mott insulator side. The unit for lattice momentum $k$ is $a^{-1}$ with $a$ the lattice constant.
}
\label{fig:momdistmott}
\end{figure}
\begin{figure}
\subfigure[ ]{
    \includegraphics[angle=0,width=0.4\linewidth]{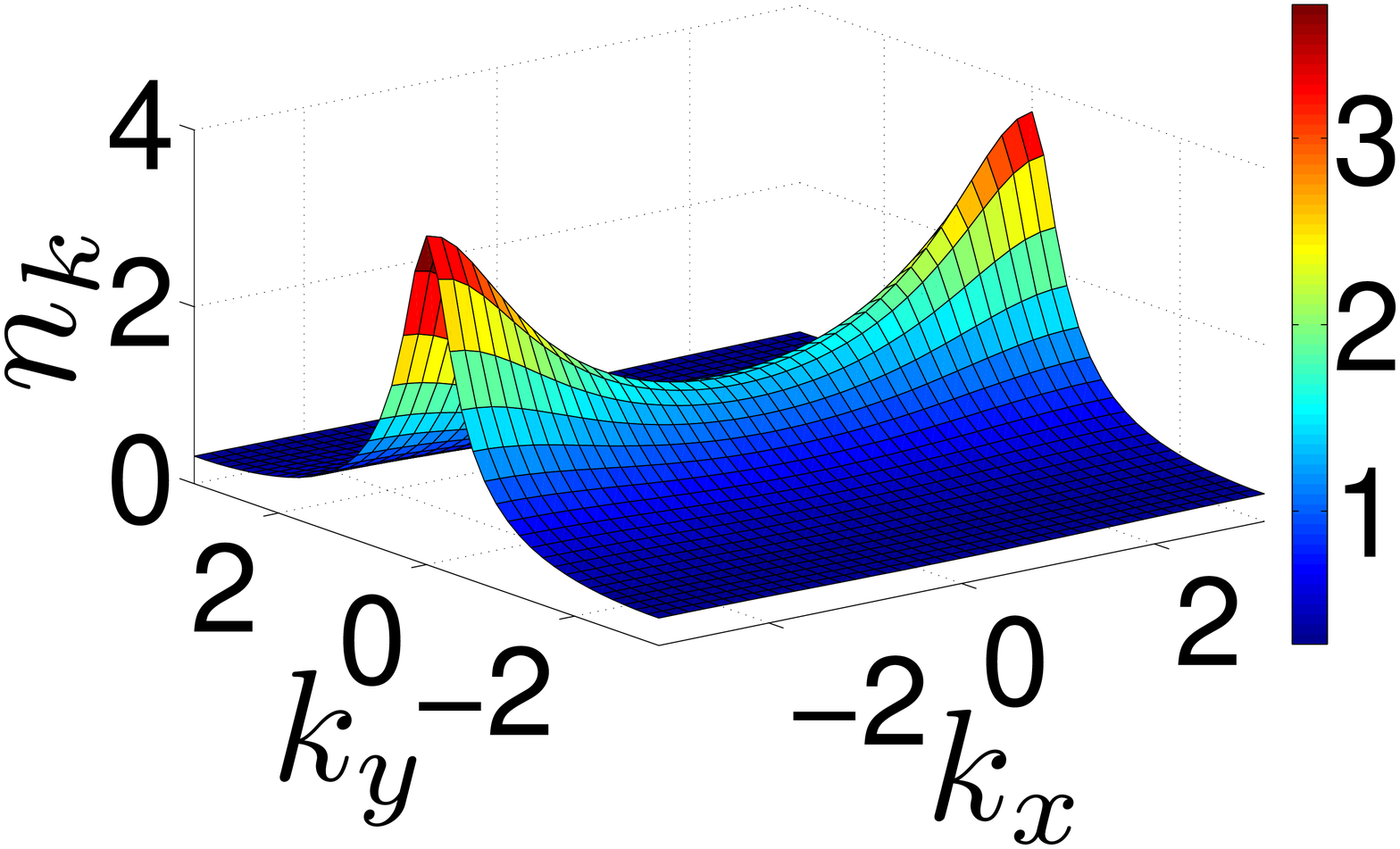}
    \label{fig:momdistmotoriginx}
}
\subfigure[ ] {
    \includegraphics[angle =0, width =0.4\linewidth]{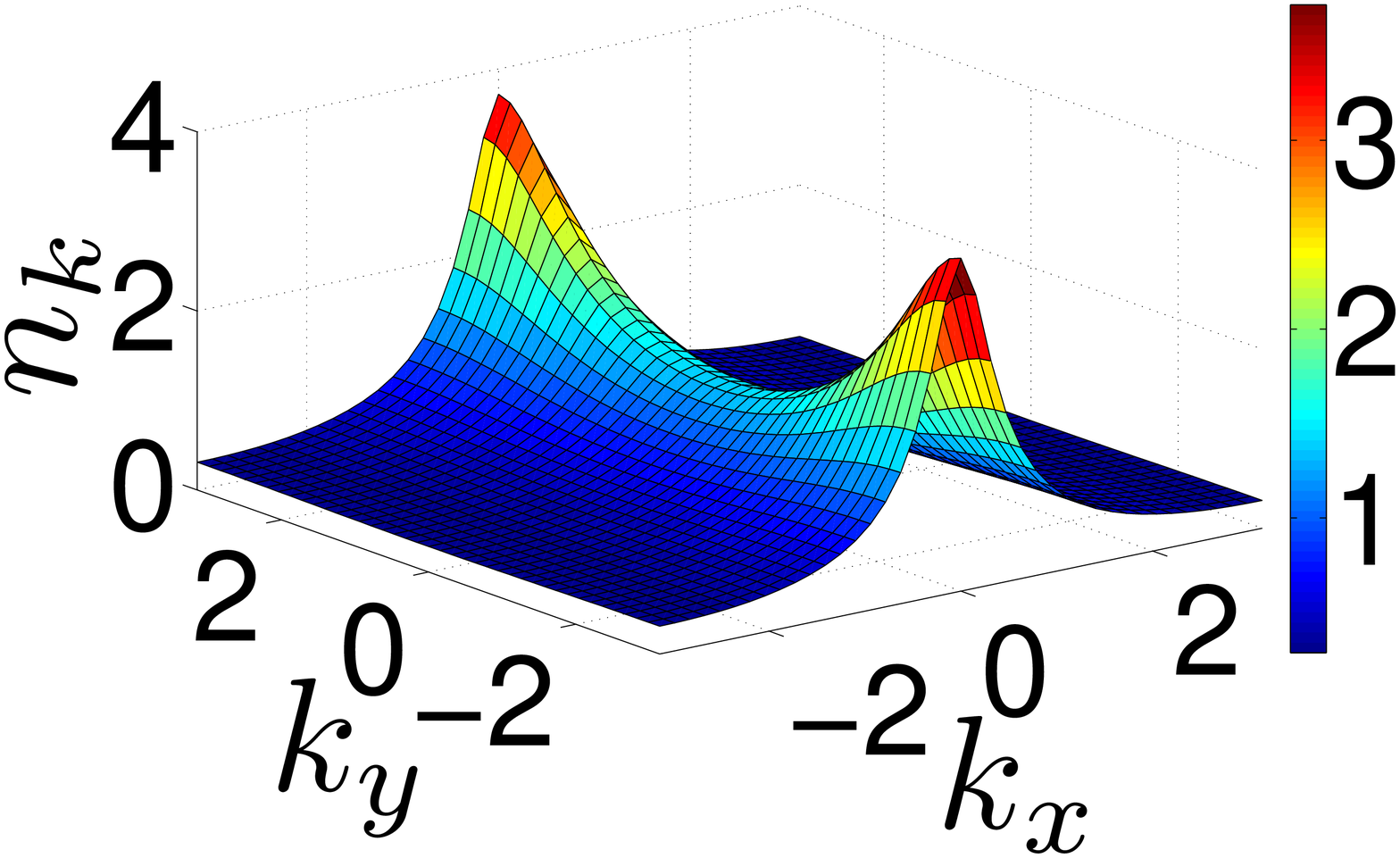}
    \label{fig:momdistmotoriginy}
}
\caption{(Color online). The momentum distribution (a) for p$_x$ band boson and (b) for p$_y$ band boson in the p-band Mott insulator phase near the Mott-superfluid phase transition point. The parameters we use are $\mu /U = 0.3$, $t/U = 0.09$ and $t_\perp =0.1 t$. The unit for lattice momentum $k$ is $a^{-1}$ with $a$ the lattice constant.
 }
\label{fig:momdistmotorigin}
\end{figure}

At zero temperature, we obtain the momentum distribution $n(\tbf{k}) \equiv \text{Tr} \langle \psi_{\sigma_1}^*(\tbf{k}) \psi_{\sigma_2} (\tbf{k}) \rangle $ of the p-band Mott insulator phase from the Green function as
$
\textstyle n(\tbf{k}) = \frac{1}{2\pi} \int d \omega e^{i\omega 0^+}
\mathscr{G}_{\sigma_1 \sigma_2} (\tbf{k}, i\omega).
$
The momentum distribution measures the spectral weight of the negative pole of Green function $[\mathscr{G} (\tbf{k}, \omega)]$ {(the negative pole refers to the real part of the energy spectrum being negative)}.%~\cite{Freericks_hubbard,Sengupta_hubbard}. 
Deep in the Mott regime, the momentum distribution is very flat (FIG.~\ref{fig:momdistmott1}). Approaching the phase transition point from Mott insulator side, a peak at zero momentum develops, and the peak diverges right at the critical point where the Mott gap closes(FIG.~\ref{fig:momdistmott2}). The continuous development of the peak of the momentum distribution at $\tbf{k} =0$ means the phase coherence ($\langle \hat{\psi}^\dag _x(\tbf{r}) \hat{\psi}_x (\tbf{r}') \rangle$ and $\langle \hat{\psi}^\dag _y(\tbf{r}) \hat{\psi}_y (\tbf{r}') \rangle$) develops continuously when increasing hopping in the Mott regime. However, we do not see the phase coherence of {$\psi_x$ and $\psi_y$ components,} i.e., $\langle \hat{\psi}_x ^\dag  (\tbf{r}) \hat{\psi}_y (\tbf{r}') \rangle$ always vanishes in the Mott
regime even when $\tbf{r}= \tbf{r} ^\prime$.
%The reason is that the global gauge symmetry $\psi_x \to \psi_x, \psi_y \to e^{i\alpha} \psi_y$ is not
%broken in the Mott regime.
To compare with experiments, we also obtain the Green function for the original bosons and get the momentum distribution of original bosons shown in FIG.~\ref{fig:momdistmotorigin}. Near phase transition point in the p-band Mott insulator phase, the momentum distribution for p$_x$ bosons shows a peak at ($\pm \pi, 0$) and the momentum distribution for p$_y$ bosons shows a peak at ($0, \pm\pi$). This offers possibilities of observing coherent matter waves from p-band Mott insulator.

\subsection{Superfluid phase}
\begin{figure}
\includegraphics[angle =0, width =0.49\linewidth]{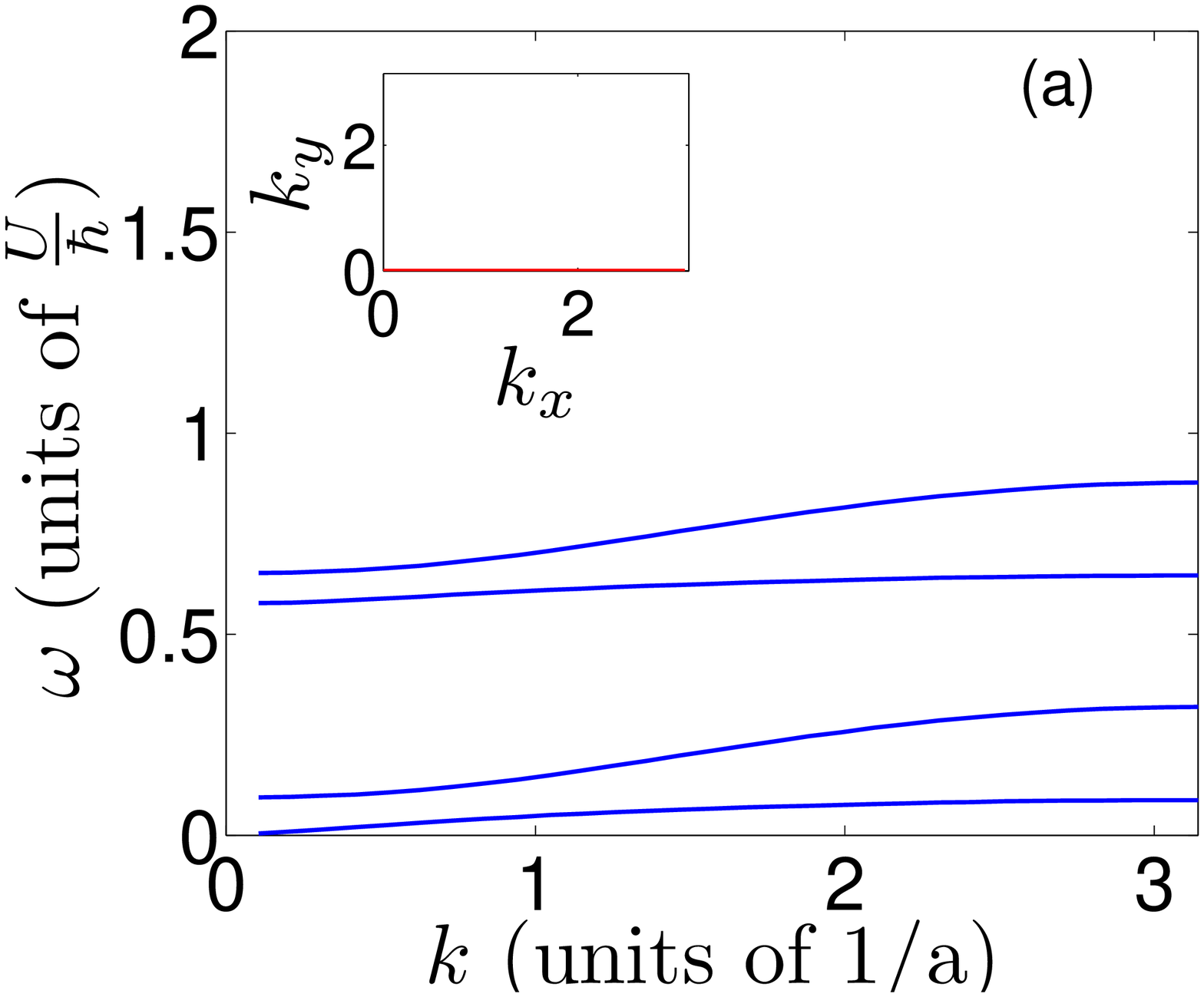}
\includegraphics[angle=0,width=0.49\linewidth]{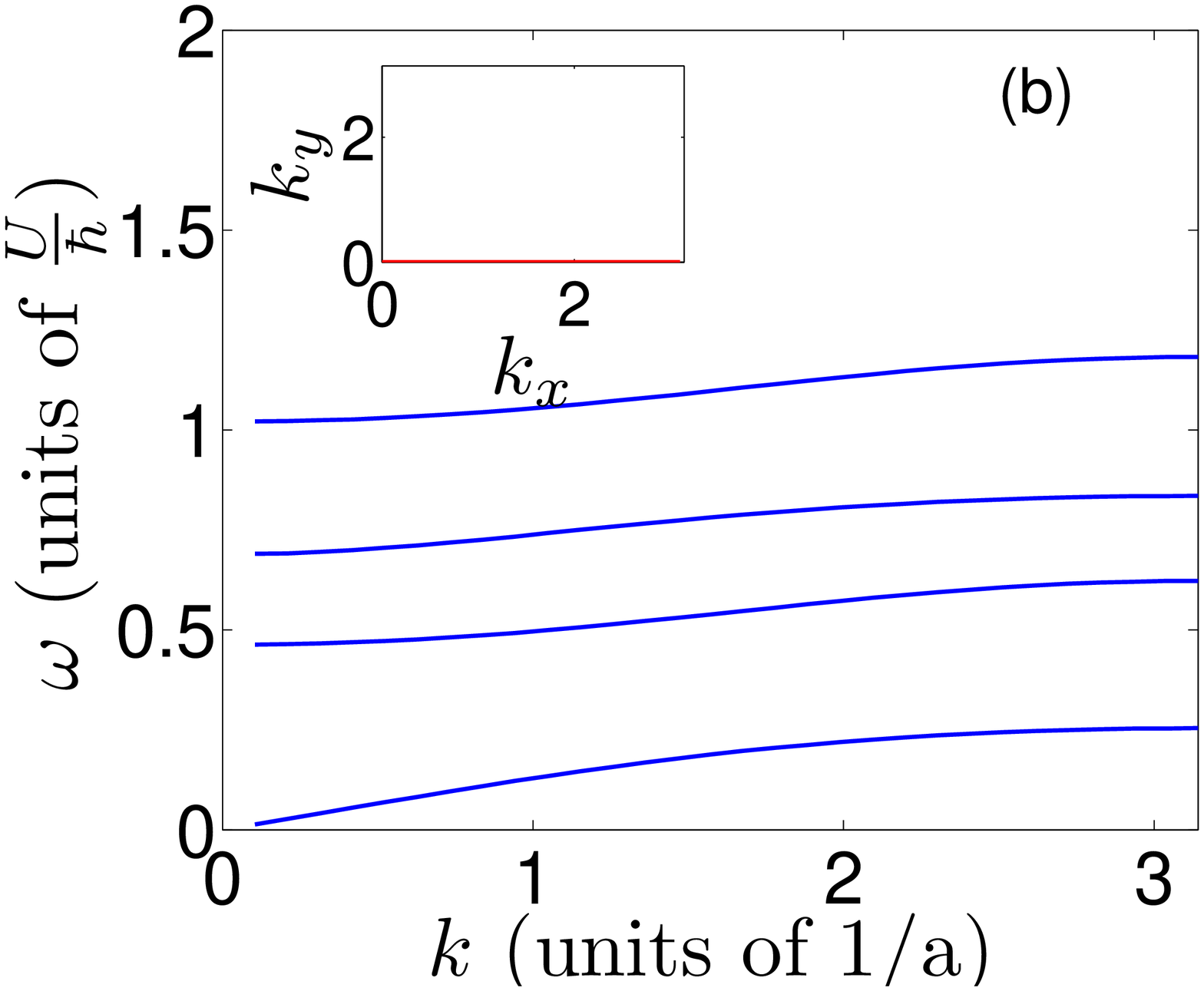}
\caption{(Color online). The Bogoliubov spectra in the TSOC superfluid phase. (a) shows the spectra along $k_x$ axis
%$\tbf{k} = k \hat{x}$ %and $\tbf{k} = k (\hat{x} + \hat{y}$)direction
near critical point with parameters
$\mu /U = 0.5$, $t/U = 0.065$, $ t_\perp =0.1 t $.
(b) shows the spectra along $k_x$ axis
%$\tbf{k} = k \hat{x}$ %and $\tbf{k} = k (\hat{x} + \hat{y}$) direction
deep in  the superfluid phase with parameters $\mu/U = 0.5, t/U = 0.1, t_\perp = 0.1t$.
}
\label{fig:sfspectra}
\end{figure}

The single particle excitation spectrum of the TSOC superfluid is calculated within Bogoliubov theory.
Solving the equation $\det [h(\tbf{k}, i\omega)] =0$, the spectrum is obtained. Deep in the TSOC superfluid phase,
there are {four branches}, one of which is gapless and linear around momentum $\tbf{k} =0$
(FIG.~\ref{fig:sfspectra}). The spectra can no longer be understood in the {same} way as in the Mott regime, since
the time reversal symmetry is broken. For momentum $k_x = \pm k_y$, the excitations have definite pseudospin,
while for generic momenta, the excitations do not carry definite pseudospin. The $\uparrow$ and $\downarrow$
components are actually mixed unless $k_x =\pm k_y$.
The physical reason for this is that the dispersion $\epsilon_{\sigma_1 \sigma_2} (\tbf{k}) $ is not
diagonal for generic momenta.

%The major difference of this superfluid phase from the s band superfluid phase is that there is one more low energy mode besides the U(1) Goldstone mode. This extra mode comes from Z$_2$ symmetry breaking, and is a mode of flipping pseudospin.
Deep in the superfluid regime, the off-diagonal term $\epsilon_{\uparrow \downarrow}$ does not affect {the U(1) phase mode} for the reason that the mode of flipping pseudospin is fully gapped (the gap is $g_{\uparrow \downarrow} n_s$) in that regime (FIG.~\ref{fig:sfspectra}(b)).
%This mode is fully gapped (the gap is roughly $g_{\uparrow \downarrow} n_s$) deep in the TSOC superfluid phase, and thus will not affect the U(1) mode in the weakly coupling regime as found in Ref.~\cite{Wu_pband}.
And the spectrum of the phase mode ($\omega_{U(1)}(\tbf{k})$) of TSOC superfluid is fully determined by $h_{11}$, $h_{12}$, $h_{21}$ and $h_{22}$ in Eq.~(\ref{hmat}), all of which are isotropic to the quadratic order of momentum $\tbf{k}$. The sound velocity defined by $\partial_\tbf{k} \omega_{U(1)}(\tbf{k})|_{\tbf{k}\to 0}$ is thus isotropic in the weak coupling regime as derived in Ref.~\cite{Wu_pband}. In the strong coupling regime, the gap of flipping pseudospin becomes smaller near the phase transition point (FIG.~\ref{fig:sfspectra}(a)), and the coupling ($\epsilon_{\uparrow \downarrow} (\tbf{k})$) between $\psi_\uparrow$ and $\psi_\downarrow$ will modify $\omega_{U(1)}(\tbf{k})$ . However, the correction to $\omega_{U(1)}(\tbf{k})$ in the limit of $\tbf{k}\to 0$ is to the order of $(t+t_\perp)^2\frac{(k_x ^2 -k_y ^2)^2}{g_{\uparrow \downarrow}n_s} $, so $\partial_{\tbf{k}} \omega_{U(1)} (\tbf{k})|_{\tbf{k}\to 0} $ is unaffected. Thus the sound velocity of TSOC superfluid is isotropic in the strong coupling regime.

%Thus the critical superfluid velocity of TSOC superfluid phase is isotropic deep in the superfluid phase, which is the same as s band superfluid phase. However, the Z$_2$ mode will modify the superfluid velocity near the critical point, because the gap of this mode ($g_{\uparrow \downarrow} n_s$) is much smaller near the critical point, and thus the coupling $h_{13}$ and $h_{24}$ are no longer negligible when calculating critical superfluid velocity near the critical point. Expanding $h_{13}$ and $h_{24}$ to quadratic order of $\tbf{k}$, one get $\frac{1}{2} (t+t_\perp) (k_x ^2 - k_y ^2)$ which is anisotropic. We calculate the critical superfluid velocity including the effect of $h_{13}$ and $h_{24}$ and find that the critical superfluid velocity (the slope of lowest branch of spectra at the $\tbf{k} =0$) is {\bf anisotropic} near critical point (FIG.~\ref{fig:sfspectra}), which is different from s band superfluidity.

\subsection{Finite temperature phase transitions of TSOC superfluid phase.}

\begin{figure}
\includegraphics[angle=0,width=1\linewidth]{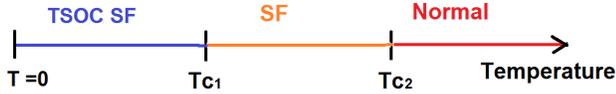}
\caption{(Color online). The schematic plot of finite temperature phase diagram in the TSOC superfluid regime. Tc$_1$ is the Ising transition temperature. Below Tc$_1$, the orbital order and the superfluid order coexist for the TSOC superfluid phase. Tc$_2$ is the KT transition temperature. For temperature above Tc$_1$ and below Tc$_2$, the orbital order no longer exists but the superfluid order still survives. Above Tc$_2$, no order exists and  the lattice bose gas is in the normal phase.
}
\label{fig:Tc}
\end{figure}

In the TSOC superfluid phase, we can neglect the temporal fluctuations of the fields. Thus we have
\be
\left[ \begin{array} {c}
\varphi_{x} (\tbf{r}) \\
\varphi_y (\tbf{r})
\end{array} \right] = \sqrt{n_s/2}
\left[ \begin{array} {c}
e^{i\theta(\tbf{r})} (-) ^{s_x (\tbf{r})}   \\
e^{i\theta(\tbf{r}) + \frac{\pi}{2} } (-) ^{s_y (\tbf{r})}
\end{array} \right],
\ee
where $s_{x/y} (\tbf{r}) = 0 \text{ or } 1$. {This substitution captures the thermal fluctuations of the TSOC superfluid phase in the strong coupling regime.}
And the energy functional describing the fluctuations of $\theta$ and $s$ fields is
\bea
&&E= -J_\theta %(-t+t_{\perp}) n_s /2
\sum_{\tbf{r}} \left[
\cos(\theta(\tbf{r}) - \theta(\tbf{r} +\hat{x})) +\cos(\theta(\tbf{r})  -\theta (\tbf{r} +\hat{y}) )
\right] \nonumber \\
&& +  \sum_{\tbf{r}}  \left [-J_1 e^{i \pi s_y (\tbf{r})} e^{i \pi s_y (\tbf{r} +\hat{y}) }
	-J_2  e^{i \pi s_y (\tbf{r})} e^{i \pi s_y (\tbf{r} +\hat{x}) } \right] \nonumber \\
&&+ \sum_{\tbf{r}}    \left [-J_1 e^{i \pi s_x (\tbf{r})} e^{i \pi s_x (\tbf{r} +\hat{x}) }
	-J_2 e^{i \pi s_x (\tbf{r})} e^{i \pi s_x (\tbf{r} +\hat{y}) } \right],
\eea
where $J_\theta = (|t|+|t_{\perp}|) n_s /2 $, $J_1 = |t| n_s /2$ and $J_2 = |t_\perp| n_s /2$. The $\theta$ part is an isotropic XY model and the $s_x$ $s_y$ parts are anisotropic Ising model.
Considering the thermal fluctuations of $\theta(\tbf{r})$, $s_x (\tbf{r})$ and $s_y (\tbf{r})$ separately, we can define two transition temperatures---the  Kosterlitz Thouless (KT) transition temperature T$_{\text{KT}}$ and the Ising transition temperature T$_{\text{Ising}}$. T$_{\text{KT}}$ is well approximated by T$_{\text{KT}} \approx \frac{\pi}{2} J_\theta $~\cite{KT}. The Ising temperature is T$_{\text{Ising}} = \frac{2}{\log (1+\sqrt{2})} J_{\text{eff}} $,  where $J_{\text{eff}}$ is determined by $\sinh(2J_1/\text{T}_\text{Ising}) \sinh (2J_2/\text{T}_\text{Ising}) = \sinh^2 (2J_\text{eff}/\text{T}_\text{Ising})$~\cite{Onsager_ising}.
%and $J_{\text{eff}}$ satisfies $\sqrt{J_1 J_2} <J_{\text{eff}} < \frac{J_1+J_2}{2}$ for our model. (Actually $J_\text{eff} \approx \frac{1}{4}  (\sqrt{J_1} +\sqrt{J_2} )^2 $ when $0.1<|\frac{J_2}{J_1}|<1$.)
For $|t_\perp| \ll |t|$, T$_{\text{Ising}} \ll$ T$_{\text{KT}}$ because $J_{\text{eff}} \to 0$ when $ t_\perp \to 0$. At temperature lower than T$_{\text{Ising}}$, the TSOC superfluid phase has an algebraic U(1) ordering and a long range Ising ordering (orbital order). At temperature higher than T$_{\text{Ising}}$, the orbital ordering disappears. Because of the existence of free Ising kinks (flips of pseudospin), half vortices with boundary condition $\oint d \theta  = \pi$ are possible, which will modify the KT transition temperature. The modified KT transition temperature is T$_\text{KT} \approx \frac{\pi}{8} J_\theta $  when $|t_\perp|\ll |t|$. The schematic finite temperature phase diagram is shown in FIG.~\ref{fig:Tc}.

\section{Conclusion}
In this paper we have developed a theory of an extended Bose-Hubbard model with p-orbital degrees of freedom, which is valid in the strong coupling regime. With this theory, we explored the single particle spectra in the Mott insulator phase and the Bogoliubov quasi-particle spectra in the TSOC superfluid phase.  We studied how the momentum distribution develops in the Mott insulator phase when increasing the inter-site hopping and found that diverging peaks rise in momentum distribution at finite momenta when the Mott gap closes. We explained the isotropy of the sound velocity of the TSOC superfluid phase. The finite temperature phase transitions of the strongly interacting TSOC superfluid phase are discussed.

\section{Acknowledgment }
We appreciate the very helpful discussions with Immanuel Bloch.  This work is supported in part
by Army Research Office (Grant No. W911NF-07-1-0293) and  DARPA OLE Program through a grant from ARO (W911NF-07-1-0464) (X.L. and W.V.L.) and Office of Naval Research (Grant No. N00014-09-1-1025A) (E.Z.).
We thank the Kavli Institute for
Theoretical Physics at UCSB for its hospitality where
this research is supported in part by National Science Foundation Grant No. PHY05-51164. X.L. is grateful for the fellowship of the A. W. Mellon Foundation.

\appendix
\section{The coefficients of the effective action}
We calculate the coefficients of the effective action in Eq.~(\ref{Seff}) by identifying the connected correlators of the
effective action and that of the original Hamiltonian (Eq.~\ref{Hor}).
The connected correlators of the original local action is calculated in the operator
representation in the occupation number basis.
\bea
&&\text{G} _\uparrow (\tbf{r}, \tau) =\frac{1}{Z_0}
 \text{Tr}[ T_\tau \hat{\psi}_\uparrow (\tbf{r},\tau) \hat{\psi}^\dag_\uparrow (\tbf{r}, 0) e^{-\beta H_0}] , \nonumber \\
&&= \frac{1}{Z_0} \sum_{n,m}  \langle n m|e^{-(\beta-\tau) H_0}   \hat{\psi}_\uparrow (\tbf{r})
e^{-\tau H_0}\hat{\psi}^\dag_\uparrow (\tbf{r}, 0)    |n m\rangle    ,\nonumber \\
&&=   \frac{1}{Z_0} \sum_{n,m} (n+1) e^{-(\beta-\tau)\epsilon(n,m) } e^{-\tau \epsilon(n+1,m)}
\eea
where
$|nm \rangle = \frac{ (\hat{\psi}_\uparrow ^\dag)^n (\hat{\psi}_\downarrow ^\dag)^m}
{\sqrt{n! m!}}|0\rangle$
are the eigenbasis
of the local interaction $H_0$ with eigenvalues
$\epsilon(n,m) = \frac{U}{2}\left((n+m)^2 -\frac{2}{3}(n+m) -\frac{1}{3}(n-m)^2\right) -\mu(n+m)$.
The fourier transform of this correlator gives
\bea
&&\mathcal{G}_\uparrow (i\omega)
\equiv \int d\tau \text{G}_\uparrow (\tau) e^{i \omega \tau},
\nonumber \\
&&= \frac{1}{Z_0}\sum_{n,m} (n+1) \left\{  \frac{ e^{-\beta \epsilon(n+1,m)}}{i\omega +\epsilon(n,m) -\epsilon(n+1,m)}
\right.\nonumber \\
 &&~~~~~~~~ \left. -\frac{e^{-\beta\epsilon(n,m)}}{i\omega +\epsilon(n,m) - \epsilon(n+1,m)} \right \}.
\eea
{In the low temperature limit ($\beta U \gg 1$), the exponential term selects out the local ground state, contributions from other states being suppressed.}
The double degenerate ground states are $|1\rangle \equiv |n_0, 0\rangle $ and
$|2\rangle \equiv|0, n_0 \rangle$, where $n_0$ is defined by
minimizing $\epsilon(n,m)$. These two ground states are related by time reversal symmetry.
Because of the double degeneracy, we further define correlators with respect to one single
ground state. $\mathcal{G}^{(1)} _\uparrow(i\omega) $ is defined corresponding to the ground state $|1\rangle$, while
$\mathcal{G}^{(2)} _\uparrow(i \omega) $ is defined corresponding to the ground state $|2 \rangle $. Thus,
\bea
\mathcal{G}^{(1)} _\uparrow (i\omega) &=& \frac{n_0}{i\omega +\epsilon(n_0-1,0) -\epsilon(n_0,0)}  \nonumber \\
&&-\frac{n_0+1}{i\omega +\epsilon(n_0,0) -\epsilon(n_0+1,0)}, \nonumber \\
\mathcal{G}^{(2)} _\uparrow(i\omega) &=& -\frac{1}{i\omega +\epsilon(0,n_0) -\epsilon(1,n_0)}, \nonumber \\
\mathcal{G}_\uparrow (i\omega) &=&\frac{\mathcal{G}^{(1)} _\uparrow +\mathcal{G}^{(2)} _\uparrow}{2}.
\label{eq:gvarphi}
\eea
It can be verified that $\mathcal{G}^{(1)} _\uparrow = \mathcal{G}^{(2)} _\downarrow$ and
$\mathcal{G}^{(2)} _\uparrow = \mathcal{G}^{(1)} _\downarrow$. Up to this point, the quadratic part of the effective action in
Eq.~(\ref{Seff}) is obtained.
In the following, we always split the correlators into two parts (labeled by superindices 1,2) according to two ground states
$|1\rangle$ and $|2\rangle$. In this paper we focus on the Mott insulator with filling $\nu =1$ for which the time reversal
symmetry is not broken. For the vortex-antivortex Mott insulator with filling $\nu >1$, the local ground state will spontaneously
choose either $|1 \rangle $ or $|2 \rangle$.  And thus one can calculate all the correlators assuming that the ground state is $|1\rangle $ or $|2 \rangle$ instead of taking the average.

The four point function is given as
\bea
&&\chi _{\sigma_1 \sigma_2} (\tau_1,\tau_2,\tau_3,0) =
\langle \psi_{\sigma_1} ( \tau_1) \psi_{\sigma_1} ^* ( \tau_2)  \psi_{\sigma_2 } (\tau_3)
\psi_{\sigma_2} ^* (0) \rangle^c _0, \nonumber \\
& = &  \frac{1}{Z_0} \text{Tr}\left[ e^{-\beta H_0} T_\tau ( e^{\tau_1 H_0}\hat{\psi}_{\sigma_1} e^{-\tau_1 H_0})
 (e^{\tau_2 H_0}\hat{\psi}_{\sigma_1} ^\dag e^{-\tau_2 H_0}) \right.\nonumber \\
&~~&~~~~~~~~~ \left. ( e^{\tau_3 H_0}\hat{\psi}_{\sigma_2 }
e^{-\tau_3 H_0})
\hat{\psi}_{\sigma_2} ^\dag \right]^c .
\eea
To calculate the coefficients $g_{\sigma_1 \sigma_2}$ in the static limit, we are only interested in the time average of
$\chi$.
\bea
\bar{\chi}_{\sigma_1 \sigma_2} &=& \int d\tau_1 d \tau_2 d \tau_3
\chi _{\sigma_1 \sigma_2} (\tau_1,\tau_2,\tau_3,0) \nonumber \\
&\equiv& \frac{\bar{\chi}^{ (1)} _{\sigma_1 \sigma_2}
+\bar{\chi}^{ (2)}_{\sigma_1 \sigma_2}}{2} .
\eea
The diagonal part is calculated as
\bea
\bar{\chi}^{(1,2)} _{\uparrow \uparrow}  &=&
 \frac{1}{Z_0} \text{Tr}_{1,2} \left[ e^{-\beta H_0} T_\tau ( e^{\tau_1 H_0}\psi_{\uparrow} e^{-\tau_1 H_0})  \right.
\nonumber \\
 &~& \left.(e^{\tau_2 H_0}\psi_{\uparrow} ^\dag e^{-\tau_2 H_0})  ( e^{\tau_3 H_0}\psi_{\uparrow } e^{-\tau_3 H_0})
\psi_{\uparrow} ^\dag \right]\nonumber \\
& -& 2\beta |\mathcal{G}_\uparrow ^{(1/2)} (0)| ^2~,
%\bar{\text{G}}^{\text{II} , (2)} _{\uparrow, \uparrow}  &=&
 %\frac{1}{Z_0} \text{Tr}_2[ e^{-\beta H_0} T_\tau ( e^{\tau_1 H_0}\psi_{\uparrow} e^{-\tau_1 H_0})
 %(e^{\tau_2 H_0}\psi_{\uparrow} ^\dag e^{-\tau_2 H_0})  ( e^{\tau_3 H_0}\psi_\uparrow e^{-\tau_3 H_0})
%\psi_\uparrow ^\dag ] - 2\beta |\mathcal{G}_\uparrow ^{(2)} (0)| ^2,
\eea
where $\text{Tr}_{1,2}$ means taking the trace with respect to the ground state $|1\rangle $ or $|2 \rangle$.
After somewhat tedious calculation we get
\bea
\bar{\chi}^{ (1)} _{\uparrow \uparrow}
&=& \frac{-4(n_0+1)(n_0+2)}{[\epsilon(n_0,0)-\epsilon(n_0+1,0)]^2 [\epsilon(n_0,0) - \epsilon(n_0+2,0)] } \nonumber \\
&+& \frac{-4(n_0 -1)n_0}{[\epsilon(n_0,0) - \epsilon(n-1,0)]^2[\epsilon(n_0,0) - \epsilon(n_0 -2,0)]} \nonumber \\
&+& \frac{-4n_0(n_0 +1)}{[ \epsilon(n_0 -1, 0) -\epsilon(n_0, 0)]^2 [ \epsilon(n_0 +1, 0)-\epsilon(n_0, 0)]} \nonumber \\
&+ & \frac{+4n_0 (n_0 +1)} {[\epsilon(n_0, 0)-\epsilon(n_0 +1,0) ]^2 [\epsilon(n_0, 0) -\epsilon(n_0 -1,0)] } \nonumber \\
&+& \frac{-4 n_0 ^2 }{ [\epsilon(n_0 -1,0)- \epsilon(n_0,0)]^3} \nonumber \\
&+& \frac{4(n_0 +1)^2} {[\epsilon(n_0,0) -\epsilon(n_0 +1,0)]^3}~~~,
\eea
\bea
\bar{\chi}^{ (2)} _{\uparrow \uparrow}
&= &  \frac{-8} {[\epsilon(0, n_0)-\epsilon(1, n_0)]^2 [\epsilon(0,n_0) - \epsilon(2,n_0)] } \nonumber \\
&+& \frac{4}{[\epsilon(0,n_0) -\epsilon(1, n_0) ]^3}~~~.
\eea
Because of time reversal symmetry other four point correlators are readily obtained by
$\bar{\chi}^{ (1)} _{\downarrow, \downarrow}
= \bar{\chi}^{(2)} _{\uparrow, \uparrow}$,
$\bar{\chi}^{ (2)} _{\downarrow, \downarrow}
= \bar{\chi}^{ (1)} _{\uparrow, \uparrow}$.

Similarly the off-diagonal part is calculated as follows
\bea
\bar{\chi}^{ (1,2)} _{\uparrow \downarrow}  &=&
 \frac{1}{Z_0} \text{Tr}_{1,2} \left[ e^{-\beta H_0} T_\tau ( e^{\tau_1 H_0}\psi_{\uparrow} e^{-\tau_1 H_0})
  \right. \nonumber \\
&~&\left. (e^{\tau_2 H_0}\psi_{\uparrow} ^\dag e^{-\tau_2 H_0})
( e^{\tau_3 H_0}\psi_{\downarrow } e^{-\tau_3 H_0})
\psi_{\downarrow} ^\dag \right ] \nonumber \\
& -& \beta \mathcal{G}_\uparrow ^{(1)} (0) \mathcal{G}_\downarrow ^{(1)} (0)  ,
 %\nonumber \\
%\bar{\text{G}}^{\text{II} , (2)} _{\downarrow, \uparrow}  &=&
 %\frac{1}{Z_0} \text{Tr}_2[ e^{-\beta H_0} T_\tau ( e^{\tau_1 H_0}\psi_{\downarrow} e^{-\tau_1 H_0})
 %(e^{\tau_2 H_0}\psi_{\downarrow} ^\dag e^{-\tau_2 H_0})  ( e^{\tau_3 H_0}\psi_{\uparrow } e^{-\tau_3 H_0})
%\psi_{\uparrow} ^\dag ] - \beta \mathcal{G}_\downarrow ^{(2)} (0) \mathcal{G}_\uparrow^{(2)} (0),
\eea
where $\text{Tr}_{1,2}$ means taking the trace with respect to the ground state $| 1 \rangle$ or $| 2 \rangle$.
After some
straightforward calculation,
\be
\textstyle \bar{\chi}^{ (1)} _{\uparrow \downarrow}
=n_0 f_0 + (n_0 +1) f_1,
\ee
with
\bea
f_0 & = &
 \frac{-1}{[\epsilon(n_0, 0) -\epsilon(n_0 -1,0)]^2 [\epsilon(n_0, 0) -\epsilon(n_0 -1, 1)]  } \nonumber \\
 & +& \frac{1} {[ \epsilon(n_0,0) -\epsilon(n_0 -1,0)]^2[ \epsilon(n_0, 0) -\epsilon(n_0,1)]}  \nonumber \\
 &+& \frac{1}{ [\epsilon(n_0,0) -\epsilon(n_0,1)]^2  [\epsilon(n_0,0) -\epsilon(n_0 -1,0)] } \nonumber \\
 &+& \frac{-1}{[\epsilon(n_0, 0) -\epsilon(n_0,1)]^2 [\epsilon(n_0,0) -\epsilon(n_0 -1,1)] } \nonumber \\
 &+&  \frac{-2  /  [ \epsilon(n_0, 0) -\epsilon(n_0  -1,1)]}
{ [\epsilon(n_0, 0) -\epsilon(n_0 -1,0)] [ \epsilon(n_0,0) -\epsilon(n_0,1)] }  ,
\eea
and
\bea
f_1 & = &  \frac{ 1}{[\epsilon(n_0, 0) -\epsilon(n_0, 1)]^2 [\epsilon(n_0,0) - \epsilon(n_0 +1,0)]} \nonumber \\
  &+ & \frac{1} { [ \epsilon(n_0, 0) -\epsilon(n_0 +1,0)]^2 [\epsilon(n_0,0) -\epsilon(n_0,1)]} \nonumber \\
 &+& \frac{-1}{[\epsilon(n_0, 0) - \epsilon(n_0,1) ]^2 [ \epsilon(n_0, 0) -\epsilon(n_0 +1, 1)]} \nonumber \\
 &+& \frac{-1 }{[\epsilon(n_0, 0) -\epsilon(n_0 +1, 0)]^2 [ \epsilon(n_0,0) -\epsilon(n_0 +1,1)] } \nonumber \\
 & + & \frac{ -2 / [\epsilon(n_0,0) - \epsilon(n_0 +1,1)] }
 {[ \epsilon(n_0, 0) -\epsilon(n_0 +1,0)][\epsilon(n_0,0) -\epsilon(n_0,1)]}.
\eea

$\bar{\chi}^{ (2)} _{\uparrow \downarrow}$ is obtained from
 $\bar{\chi}^{ (1)} _{\uparrow \downarrow}$ with $\epsilon(n,m)$ substituted
by $\epsilon(m,n)$.
For $n_0 >1$, one can safely let
$\bar{\chi}^{ (2)} _{\uparrow \downarrow} = \bar{\chi}^{ (1)} _{\uparrow \downarrow} $
 and thus
$\bar{\chi} _{\uparrow \downarrow} =\bar{\chi}^{ (1)} _{\uparrow \downarrow}$
because $\epsilon(n,m) = \epsilon(m,n)$.
However for $n_0 =1$, both $\bar{\chi}^{ (2)} _{\uparrow \downarrow}$ and
$ \bar{\chi}^{ (1)} _{\uparrow \downarrow} $ are singular because
$\epsilon(1,0) = \epsilon(0,1)$; while the average of these two
is finite by taking the proper limit $\epsilon(1,0) = \epsilon(0,1) + 0^+$.

Up to this point, we have obtained the four point correlators from the original Hamiltonian in Eq.~(\ref{Hor}). In order to calculate
$g_{\sigma_1 \sigma_2}$, we still need the time average of the four point correlator defined by the effective action in
Eq.~(\ref{Seff}).
Since the local ground state of this theory is not unique, one should not
naively apply the Feynman rules of the usual $\phi^4$ field theory.
%cannot blindly apply Feynman rules.
Similar to the approach
we used above, we calculate the correlators on each ground state and then take the average. Thus we have
\bea
&&\bar{\chi }_{\sigma_1 \sigma_2} = \int d\tau_1 d \tau_2 d \tau_3
\chi _{\sigma_1 \sigma_2} (\tau_1,\tau_2,\tau_3,0) \nonumber \\
&& =  \frac{\bar{\chi}^{ (1)} _{\sigma_1 \sigma_2}
+\bar{\chi}^{ (2)}_{\sigma_1 \sigma_2}}{2} \nonumber \\
&&=  - g_{\sigma_1 \sigma_2} \left\{ [ (\mathcal{G}^{(1)} _{\sigma_1}(0) \mathcal{G}^{(1)} _{\sigma_2}(0))^2
+(\mathcal{G}^{(1)} _{\sigma_1})^4 \delta_{\sigma_1, \sigma_2}  ] \right. \nonumber \\
&&+ \left. [ (\mathcal{G}^{(2)}  _{\sigma_1}(0) \mathcal{G}^{(2)} _{\sigma_2}(0))^2
+(\mathcal{G}^{(2)} _{\sigma_1})^4 \delta_{\sigma_1, \sigma_2} ]\right \}/2 .
\eea
Then $g_{\sigma_1 \sigma_2}$ is obtained,
\bea
g_{\sigma_1 \sigma_2} &=& -2 \bar{\chi }_{\sigma_1 \sigma_2}/
 \left\{ [ (\mathcal{G}^{(1)} _{\sigma_1}(0) \mathcal{G}^{(1)} _{\sigma_2}(0))^2
+(\mathcal{G}^{(1)} _{\sigma_1})^4 \delta_{\sigma_1, \sigma_2}  ] \right. \nonumber \\
&+& \left. [ (\mathcal{G}^{(2)}  _{\sigma_1}(0) \mathcal{G}^{(2)} _{\sigma_2}(0))^2
+(\mathcal{G}^{(2)} _{\sigma_1})^4 \delta_{\sigma_1, \sigma_2} ]\right \} .
\eea
We verify that these coefficients satisfy $g_{\uparrow \downarrow} > g_{\uparrow \uparrow}$.

\section{Validity of the double Hubbard-Stratonovich transformation} 
The connected correlators of the original boson fields $\psi_\sigma$ can be obtained from the generating functional
\bea
\textstyle Z[J^*, J] \textstyle  \equiv 
e^ { \textstyle -\psi^* _\sigma T _{\sigma \sigma'} \psi _{\sigma'} -S_0 [\psi ^*, \psi] +[(J| \psi) +c.c.]  }.
\eea
For example,
\bea
\textstyle \langle \psi_{\sigma} (x) \psi_{\sigma'} ^* (x') \rangle_{S [\psi ^*, \psi]} 
\textstyle = \lim_{J\to 0} \frac{\delta \log Z[J^*, J]}{\delta J_\sigma ^* (x) \delta J_{\sigma'} (x')} ,
\eea
where $x \equiv (\vec{r}, \tau)$ and $\langle$\ldots$\rangle_{S [\psi ^*, \psi]}$ means an average defined by the action
$S [\psi ^*, \psi] = \psi^* _\sigma T _{\sigma \sigma'} \psi _{\sigma'} +S_0 [\psi ^*, \psi]$. 
%{\color{red} For simplicity, the summation over $\vec{r}$ and the integral over $\tau$ is not shown explicitly in this appendix.} 
Introducing the first Hubbard-Stratonovich transformation,
\bea
\textstyle Z[J^ *, J] &&= \textstyle \int D[\psi^*, \psi; \phi ^*, \phi]
\exp \left \{
\phi_\sigma ^* T ^{-1} _{\sigma \sigma'} \phi_{\sigma '} -S_0 [\psi_\sigma ^*, \psi_\sigma] \right. \nonumber \\
&&\textstyle \left.+[(\phi|\psi) +c.c.] +[(J|\psi) +c.c.]
\right \} .
\eea
After a shift $\phi_\sigma \to \phi_\sigma -J_\sigma$, we get
\bea
\textstyle Z[J^*, J] %= && \textstyle \int D[\psi ^*, \psi, \phi ^* , \phi]
%\exp \left \{ \textstyle
%(\phi_\sigma- J_\sigma ) ^* T ^{-1} _{\sigma \sigma'} (\phi_{\sigma '} - J_{\sigma '} ) +[(\phi-J|\psi) +c.c.] + [(J|\psi) +c.c.]
%\right \} \nonumber \\
%&&\times e^{-S_0 [\psi_\sigma ^*, \psi_\sigma]}  , \\
&& \textstyle = \int D[\psi ^*, \psi, \phi ^* , \phi]
e^{  [(\phi|\psi) +c.c.]-S_0 [\psi_\sigma ^*, \psi_\sigma]} \nonumber \\
&& \times \exp \left \{ \textstyle 
(\phi_\sigma- J_\sigma ) ^* T ^{-1} _{\sigma \sigma'} (\phi_{\sigma '} - J_{\sigma '} )
\right \} .
\eea
After integrating out the $\psi_\sigma$ fields, we get
\bea
\textstyle Z[J^*, J] &&= \textstyle Z_0 \int D[\phi ^* , \phi]
\exp \left \{
(\phi_\sigma- J_\sigma ) ^* T ^{-1} _{\sigma \sigma'} (\phi_{\sigma '} - J_{\sigma '} ) \right.\nonumber \\
\textstyle && \left . + W [\phi_\sigma ^*, \phi_\sigma] \right \} .
\eea
Now we introduce the second Hubbard-Stratonovich transformation
\bea
&&\textstyle Z[J^*, J] \nonumber \\ 
%&&= Z_0 \int D[\varphi^*, \varphi; \phi^*, \phi]
%\exp \left \{
%-\varphi_\sigma ^* T_{\sigma \sigma '} \varphi_{\sigma '} - [(\varphi| \phi - J )+c.c.] + W[\phi_\sigma ^*, \phi_\sigma]
%\right \}, \nonumber \\
&& \textstyle = Z_0 \int D[\varphi^*, \varphi; \phi^*, \phi]
\exp \left \{ \textstyle 
-\varphi_\sigma ^* T_{\sigma \sigma '} \varphi_{\sigma '} + [(\varphi| J )+c.c.] \right \} \nonumber \\
&& \textstyle  \times \exp \left \{-[(\varphi|\phi) +c.c.] + W[\phi_\sigma ^*, \phi_\sigma] \right \} .
\eea
Integrating out the $\phi_\sigma$ fields, we get
\bea
\textstyle Z[J^*, J] && \textstyle =Z_0 \int D[\varphi^*, \varphi]
\exp \textstyle \left \{ \textstyle
-\varphi_\sigma ^* T_{\sigma \sigma '} \varphi_{\sigma '}  +
\textstyle \tilde{W} [\varphi_\sigma ^*, \varphi_\sigma] \right. \nonumber \\
&& \textstyle  \left. + [(\varphi| J )+c.c.]\right \} .
\eea
The generating functional for $\varphi_\sigma$ fields is equal to the generating functional of $\psi_\sigma$ (up to a constant independent of the sources $J$). And this proves the connected correlators of $\varphi_\sigma$ are the same as that of $\psi_\sigma$, although the action of $\varphi_\sigma$ is very different from that of $\psi_\sigma$. For example,
\bea
%&&\langle \psi_{\sigma} (x) \psi_{\sigma'} ^* (x') \rangle_{S [\psi ^*, \psi]}
%= \lim _{J\to 0} \frac{\delta Z[J^*, J]}{\delta J_\sigma ^* (x) \delta J_{\sigma'} (x')}, \nonumber \\
&&\textstyle \langle \varphi_{\sigma} (x) \varphi_{\sigma'} ^* (x') \rangle_{S_{\text{eff}} [\varphi ^*, \varphi]}
= \textstyle \lim _{J\to 0} \frac{\delta \log Z[J^*, J]}{\delta J_\sigma ^* (x) \delta J_{\sigma'} (x')}, \nonumber \\
&&= \langle \psi_{\sigma} (x) \psi_{\sigma'} ^* (x') \rangle _{S [\psi ^*, \psi]}. 
%\langle \varphi_{\sigma} (x) \varphi_{\sigma'} ^* (x') \rangle_{S [\varphi ^*, \varphi]},
\eea
where $\langle$\ldots$\rangle_{S_\text{eff} [\varphi ^*, \varphi]}$ means an average defined by the action
$S_{\text{eff}} [\varphi ^*, \varphi] = \varphi_\sigma ^* T_{\sigma \sigma '} \varphi_{\sigma '} -\tilde{W} [\varphi_\sigma ^*, \varphi_\sigma]$.

\bibliography{pband}
\bibliographystyle{apsrev}
\end{document}